\documentclass[11pt]{aastex63}
\usepackage{graphicx}
\usepackage{amsmath}

\newcommand{\mysoft}[1]{{\texttt{#1}}}

\def\Msolar{\ifmmode {\rm M_{\odot}}\else $\rm M_{\odot}$\fi}
\def\Mearth{\ifmmode {\rm M_{\oplus}}\else $\rm M_{\oplus}$\fi}
\def\Rearth{\ifmmode {\rm R_{\oplus}}\else $\rm R_{\oplus}$\fi}
\def\micron{\ifmmode {\mu{\rm{m}}}\else $\mu$m\fi}

\def\unitobs{{\hat{e}_\text{obs}}}
\def\ain{{a_\text{in}}}
\def\ainmin{{a_\text{min}}}
\def\aoutmax{{a_\text{max}}}
\def\aout{{a_\text{out}}}
\def\Sin{{\Sigma_\text{in}}}
\def\inc{\imath}

\def\tflare{t_\text{flare}}

\def\nflares{N_\text{f}}
\def\dV{{\Delta}{V}}
\def\Flux{{\cal F}}
\def\Fluxnorm{\tilde{\Flux}}

\def\Fflare{F_\text{flare}}
\def\Fecho{F_\text{echo}}
\def\Fechoi{F_\text{echo}[i]}
\def\Fmulti{F_\text{multi}}
\def\Fmultii{F_\text{multi}[i]}
\def\rmin{{r_\text{min}}}
\def\rmax{{r_\text{max}}}
\def\rp{{r_p}}
\def\Ap{{A_p}}
\def\massp{{m_p}} 
\def\rhop{\rho}
\def\rratfacexpanded{{\left(\frac{\rmax}{\rmin}\right)^{\!1/2}}}
\def\rratfac{{X}}

\def\sfracp{f_p}
\def\mdisk{{M_\text{disk}}} 
\def\mfit{{\cal M}} 
\def\mlim{{\cal M}_\text{lim}} 
\def\phasefn{\Phi}

\begin{document}

\vspace*{-1.5in}

\title{Seeking echoes of circumstellar disks in Kepler light curves}

\author{Benjamin C. Bromley}
\affil{Department of Physics \& Astronomy, University of Utah, 
\\ 115 S 1400 E, Rm 201, Salt Lake City, UT 84112}
\email{bromley@physics.utah.edu}

\author{Austin Leonard}
\affil{Nanohmics, Inc., 6201 E Oltorf St., Ste 400, Austin, TX 78741}
\email{aleonard@nanohmics.com}

\author{Amanda Quintanilla}
\affil{Nanohmics, Inc., 6201 E Oltorf St., Ste 400, Austin, TX 78741}
\email{aquintanilla@nanohmics.com}

\author{Austin J. King}
\affil{Department of Physics \& Astronomy, University of Utah, 
\\ 115 S 1400 E, Rm 201, Salt Lake City, UT 84112}
\email{austin.king@utah.edu}

\author{Chris Mann}
\affil{Nanohmics, Inc., 6201 E Oltorf St., Ste 400, Austin, TX 78741} 
\email{cmann@nanohmics.com}

\author{Scott J. Kenyon}
\affil{Smithsonian Astrophysical Observatory,
\\ 60 Garden St., Cambridge, MA 02138}
\email{skenyon@cfa.harvard.edu}

\begin{abstract}
Light echoes of flares on active stars offer the opportunity for direct detection of circumstellar dust. We revisit the problem of identifying faint echoes in post-flare light curves, focusing on debris disks from on-going planet formation. Starting with simulations, we develop an algorithm for estimating the radial extent and total mass from disk echo profiles. We apply this algorithm to light curves from over 2,100 stars observed by NASA's Kepler mission, selected for multiple, short-lived flares in either the long-cadence or short-cadence data sets. While flux uncertainties in light curves from individual stars preclude useful mass limits on circumstellar disks, catalog-averaged light curves yield constraints on disk mass that are comparable to estimates from known debris disks. The average mass in micron- to millimeter-sized dust around the Kepler stars cannot exceed 10\% of an Earth mass in exo-Kuiper belts or 10\%\ of a Lunar mass in the terrestrial zone. We group stars according to IR excess, based on WISE W1-W3 color, as an indicator for the presence of circumstellar dust. The mass limits are greater for stars with strong IR excess, a hint that echoes are lurking not far beneath the noise in post-flare light curves. With increased sensitivity, echo detection will let time-domain astronomy complement spectroscopic and direct-imaging studies in mapping how, when, and where planets form.
\end{abstract}


\section{Introduction} 
\label{intro}

The light from a star shines brightly on the planetary system it hosts. Yet the light coming directly from a planet, a swarm of comets, or a  dusty circumstellar disk may be difficult to measure because it is so faint compared to the starlight. In some cases, astronomers can get lucky, as when a planet orbits far from its host and can be spatially resolved \citep[e.g.,][]{marois2008, macintosh2015}. In other systems, starlight is reprocessed by planetary dust so that it glows in a waveband where the stellar flux is low. Then, a distinct ``infrared excess" in the stellar spectrum appears from the blackbody radiation of debris at an equilibrium temperature that is significantly lower than that of the surface of the stellar host \citep[e.g.,][]{aumann1984, walker1988, beichman2006, bryden2009}.

When a star has a time-varying luminosity there is another avenue for identifying light that originates directly from nearby planetary material. The idea is to detect echoes in the star's light curve, a strategy that has been suggested in a variety of astrophysical settings \citep[e.g.,][]{rest2012}.  Echoes of supernova explosions  \citep[e.g.,][]{crotts1989, sugerman2003} and reverberation mapping of disks around stars \citep{meng2016} and black holes \citep[e.g.,][]{horne1991, peterson2004, grier2019} are examples.

A vein of research has explored the use of echoes in flare-star light curves to detect planets \citep{argyle1974, matloff1976, bromley1992, clark2009, mann2017, sparks2018}.  \citet{mann2018} showed that multiple observations of faint echoes from planets can potentially reveal not only their presence, but a complete set of orbital elements as well. However, the operative word is faint: the starlight reflected by the Earth around the Sun is less than one part in a billion. Prospects are better for a hot jupiter in close orbit around its stellar host, but even then, the detection of an echo is challenging.
  
\citet{gaidos1994} noted that the situation is significantly different when mass is distributed in a disk. The reason is that even a small object like the Moon, if broken up into micron-sized grains, has a comparatively large surface area, capable of blotting out the central star altogether. In the scenario considered by \citet{gaidos1994} \citep[see also][]{sugerman2003}, the disk is optically thick and intercepts a good fraction of the starlight as determined by the disk geometry. The reflected starlight in this case can be a few percent of the quiescent light, so that the echo from a strong flare may be more easily resolved in the light curve following a powerful flare. However, a post-flare light curve commonly features more stellar activity than before the flare, so multiple measurements are required to help rule out microflare `aftershocks'.  
  
Bringing the potential of echo detection to fruition requires stars that offer repeated, short, powerful outbursts. Active late-type stars have up to dozens of flares per day \citep[e.g.,][]{pettersen1986, hawley2014,  balona2015, davenport2016, yang2017}. The contrast between the flare emission and the quiescent starlight can be high, exceeding $10^{4}$--$10^{5}$ in select wavebands \citep[e.g.,][]{schmidt2014, schmidt2016}. While such powerful flares may destroy close-in dust, they illuminate a large swath of the planet-forming region around these stars. Thus, the strong contrast, high flare frequency, and the prevalence of late-type stars \citep{ledrew2001} make them excellent candidates for echo detection.

The flaring activity of stars on the main sequence is highest when they are young, with ages less than $\sim 300$~Myr \citep[e,g.,][]{Zhang2019}. An important coincidence is that circumstellar debris disks --- potential sources of echoes --- are more prevalent within the first gigayear of a star's lifetime \citep[e.g.,][]{su2006, carpenter2009, eiroa2013, sibthorpe2018}. Debris disks emerge as growing planets gravitationally stir small objects to high relative speeds \citep{ida1992}; collisions between these small bodies cause them to shatter into yet more numerous smaller objects that collide and fragment even more vigorously \citep[e.g.,][]{weidenschilling1997}. This collisional cascade produces the debris disks observed around stars with on-going planet formation \citep[e.g.,][]{kb2004}. 

Additionally, growing protoplanets with radii of 100--1000~km can be numerous around stars where planets are forming. Giant collisions between pairs of these objects can produce copious amounts of dust \citep{kb2005}. While short-lived, such events may be spectacular, producing narrow rings of debris and offering the possibility of echoes that are detectable with a time-domain surveyor like NASA's Kepler satellite.

The Kepler mission \citep{Borucki1996, borucki2010} observed around 150,000 stars for a period of over four years. The flux sensitivity is roughly one part in 10,000; the Kepler Data Characteristics Handbook \citet{keplerhandbook2016} provides technical details.  The mission produced data at long cadence (29.42 minutes) and short cadence (58.85 seconds) in the form of full-frame images, target pixel files, and light curves  \citep{Koch1998, gilliland2010}.  The targets observed by Kepler are primarily F, G, and K stars, along with a smaller fraction of M dwarfs \citep{Mulders2015}. Data from these sources, which include thousands of active stars \citep[e.g.][]{balona2015}, offer a unique opportunity for identifying flare stars and placing constraints on flare echoes.

Pre-main sequence stars, including T~Tauri stars, are chromospherically active and produce a large number of flares \citep{Navascus2003}. Furthermore, virtually all T~Tauri stars have protoplanetary disks \citep{osterloh1995, ribas2014, richert2018}, making them excellent candidates for echo detection \citep{meng2016}. 
While echoes are likely promising probes of these young systems, the focus of this work is on Kepler data and the potential for detecting unresolved debris disks around main sequence stars.

In this paper, we explore in detail the possibility of detecting echoes from debris disks. In \S\ref{sec:bg} we give an overview of disk properties and light scattering by circumstellar dust. In \S\ref{sec:echoes}, we examine the nature of flare echoes from optically thin disks, and introduce a strategy for echo detection and disk parameter estimation. Then, we apply our method to roughly 100,000 light curves from over 2,000 active stars observed by the Kepler satellite (\S\ref{sec:kep}). Finally, we discuss our results in \S\ref{sec:discuss} and conclude in \S\ref{sec:conclude}.

\section{background}\label{sec:bg}

In a previous effort, we investigated the theoretical feasibility of detecting exoplanets from their light-curve echoes.  The distinct advantage of seeking light echoes from dusty disks as opposed to a planet amounts to surface area. For example, an object similar to the Moon, with density of $\rhop = 3$~g/cm$^3$ and total mass $M \approx 0.01$~M$_\oplus$, at a distance $a=1$~au from its host star, has an illuminated disk of area of 0.07 times that of the Earth ($A_\oplus$), enabling it to intercept only a small fraction, $f = 3.2\times 10^{-11}$, of the light from its stellar host. If this same planetary mass were in the form of many smaller bodies with radius $\rp$, the surface area would be much larger:
\begin{equation}
A = \frac{3 M}{4 \rhop \rp} \approx 1.2\times 10^{11} \left[\frac{M}{0.015\ \text{M}_\oplus}\right] \left[\frac{\rhop}{2~\text{g/cm}^3}\right]^{-1} \left[\frac{\rp}{1~\text{$\mu$m}}\right]^{-1} \times A_\oplus.
\end{equation}
A swarm of these particles, if distributed uniformly on a sphere of radius 1~au, centered on the host star, would paint the sphere 50 times over. Even at a distance of 10~au, a spherical shell of dust could reflect over 50\%\ of the starlight.

Although circumstellar dust in real astrophysical systems is often confined to geometrically thin disks or rings, this simple illustration suggests that dust may be extraordinarily bright when compared to the reflected light from a planet. If present in even modest amounts, dusty rings or disks have excellent potential as sources of light echoes.

\subsection{Dust properties}\label{subsec:dust}

Circumstellar dust around active, Sun-like stars comes from collisions between planetary bodies.  Theoretical arguments  \citep{dohnanyi1969, campobagatin1994, wyatt2002, kb2004, pan2005, dominik2003, krivov2006, thebault2007, kb2010, pan2012, gaspar2012, kb2017} and observations \citep[e.g.,][]{su2009, takasawa2011, su2015, MacGregor2016, MacGregor2017, marshall2017, hughes2018} indicate that the distribution of particle radii $r$ in collisional debris approximately follows a power-law, $dn/dr \sim r^{-q}$, where $q \sim 3$--4. These extended distributions arise in a steady, collisional cascade, wherein large bodies collide to produce small objects, which in turn collide to make yet smaller ones. Stellar radiation and wind clear out the smallest objects at the end of the cascade; the typical ``blowout radius'' is $\rmin \sim 1$~\micron, although some observations favor larger minimum sizes \citep[and references therein]{pawellek2015}.  

Even though astrophysical dust grains have a wide range of sizes, the steep fall-off of the differential size distribution with radius means that the average particle size and mean cross-sectional area per particle depend only on the minimum particle size:
 \begin{eqnarray}\label{eq:rp}
     \ & {r}_p \equiv \left<r\right> \ = \ \frac{q-1}{q-2}\ \rmin
     \ = \   
     \frac{5}{3} \times \rmin
     & \quad\quad (\text{mean size}; q = 3.5, \rmin \ll \rmax), \\
     \label{eq:ap}
    \ & A_p \equiv \left<\pi r^2\right> \ = \ \frac{q-1}{q-3}\ \pi\rmin^2 
     \ =  \ 
     5 \times \pi \rmin^2 
     & \quad\quad (\text{mean particle area}),
\end{eqnarray}     
where we have assumed that the radius of the largest particles in the distribution, $\rmax$, is much bigger than the blowout radius $\rmin$.  With $q=3.5$, consistent with both theory and observations, the particles that contribute most to the reflecting area of a collection of dust are small: Over 96\%\ of the total cross-sectional area comes from dust grains with sizes that range from $\rmin$ to $1000\rmin$. Thus, in astrophysical disks, \micron- to mm-sized dust generate much, if not virtually all, of the reflected light, while meter-size or larger objects escape detection. This general conclusion and the results that follow are not strongly sensitive to the specific choice of $q$ within the range allowed by observation ($3 < q < 4$).

Compared to the cross-sectional area, the mean mass per particle has a stronger dependence on $\rmax$, the radius of the largest particles in the distribution. In debris disks, where ongoing planet formation is driving a collisional cascade, or in the event of a single catastrophic collision between planets, $\rmax$ can be hundreds of kilometers \citep[e.g.,][respectively]{pan2012, leinhardt2012}. Formally,  
 \begin{eqnarray}\nonumber
     \massp & \equiv & \left<\frac{4}{3}\pi \rhop r^3\right> \ = \ \frac{4}{3} \pi \rhop \ \frac{q-1}{4-q}\ \rmax^{4-q}\rmin^{q-1} \\
     & = &   \label{eq:mp}
     \ 5 \left(\frac{\rmax}{\rmin}\right)^{\!1/2} \times \frac{4}{3}\pi \rhop \rmin^3
      \ \ \ \ \ \ \ \ \ \ \ 
      (\text{mean particle mass}),
\end{eqnarray}
where $\rhop$ is the bulk mass density of dust particles, assumed to be in the range of roughly 1.5~g/cm$^3$ (icy material) to 2.5~g/cm$^3$ (for astrophysical silicates); we adopt a fiducial value of $\rhop = 2$~g/cm$^3$. In this equation, with $q=3.5$, the mean mass per particle depends significantly on the maximum particle size ($\massp\sim \sqrt{\rmax}$). 

To quantify the dependence of the mass in Equation~(\ref{eq:mp}) on particle sizes, we define
\begin{equation}\label{eq:rratfac}
    \rratfac \equiv \rratfacexpanded.
\end{equation}
The mean mass per particle is five times $\rratfac$ in units of the mass of the smallest dust grains. If a disk were composed of mostly small particles, 1~\micron--1~mm, then $\rratfac \approx 32$. If there were an on-going collisional cascade in a disk that included kilometer-sized then $\rratfac \sim 3\times 10^4$, in which case those small bodies that account for more than 95\%\ of the reflecting surface area of the particle distribution contain only 0.1\%\ of the mass. 
\subsection{Scattering by dust}\label{subsec:scatter}

An important ingredient in the production of a light echo by a dusty disk is how individual dust particles scatter incident radiation. Typical grain sizes are roughly microns (Eq.~(\ref{eq:rp})), comparable to the wavelength of radiation from a Sun-like star; $\lambda \sim 0.1$-1~$\mu$m, for ultraviolent (UV) to infrared (IR) light. In this regime, a phase function, $\phasefn$, describes the fraction of light scattered at some angle relative to the incoming radiation. While Mie theory captures the physics of scattering for ideal scatterers, providing a quantitative prediction for the phase function, we adopt the \citet{draine2003} parameterization, which is tuned for astrophysical dust: 
\begin{equation}\label{eq:phasefn}
    \phasefn(\theta) = 
    \frac{1}{4\pi}\ \frac{1-g^2}{1+(1+2g^2)\nu/3}\ \frac{1+\nu\cos^2(\theta)}{\left[1+g^2-2g\cos(\theta)\right]^{3/2}},
\end{equation}
where $\theta$ is the scattering angle ($\theta=0$ is undeflected light, while $\theta = \pi$ is back-scattered), $g = \langle\cos(\theta)\rangle$ is the mean scattering anisotropy ($g=0,1,-1$ for isotropic, forward and back scattering, respectively), and $\nu$ is connected to scattering physics ($\nu = 1$ for Rayleigh scattering, and $\nu = 0$ yields the Henyey-Greenstein phase function for larger dust grains; \citealt{henyey1941}). We choose values of $g = 0.429$ and $\nu = 0.114$, guided by observed astrophysical dust properties in optical wavebands \citep[see][Fig.~6 therein]{draine2003}. 

The phase function adopted here is normalized such that integration over all scattering angles gives unity. This choice, along with an adopted albedo, $\alpha$, enables us to quantify the fraction of starlight that is scattered given a particle's distance from the host star and the scattering angle relative to some far-away observer. For the polydisperse dust described above ($q=3.5$), the mean fraction of scattered starlight is
\begin{equation}\label{eq:fp}
    \sfracp(a,\theta) = \alpha \frac{\Ap}{4\pi a^2} \cdot 4\pi \phasefn(\theta) 
    =  \frac{5 \pi \alpha \rmin^2}{a^2} \phasefn(\theta).
\end{equation}
The normalization of the phase function is important to the constant factors in this equation; for isotropic scattering, $\phasefn = 1/4\pi$. The total reflected light from a spatially extended collection of dust is the sum of all such contributions from all regions of the disk.

\subsection{Dust reflectivity, thermal radiation and sublimation}\label{subsec:sublimation}

The reflectivity of astrophysical dust composed of ices or silicates is high, with albedo $\alpha \gtrsim 0.5$--1, at optical wavelengths \citep[e.g.,][]{ysard2018}. In the theoretical models adopted here, we assume that dust grains are perfect scatterers ($\alpha = 1$). Even when this approximation is realistic \citep[cf.][]{krist2010}, dust may absorb radiation in other wavebands. The overall absorption of starlight by individual grains leads to heating, which in turn causes the dust to re-radiate thermal photons. When a star is in quiescence, dust achieves thermal equilibrium at a temperature $T_\text{eq}$ that depends on its material properties and the stellar flux at its orbital distance, $a$, from the star; $T_\text{eq} \sim a^{-1/2}$.  At some small distance, the equilibrium temperature exceeds the sublimation temperature of the grains. This distance defines the sublimation zone, within which grains are destroyed \citep[e.g.,][]{kobayashi2011}. For icy particles around the Sun, this limit is the ``snow line'' at roughly 3~au \citep{hayashi1981}. For silicates, the analogous ``sand line'' is within 0.1~au.

Around an active star, strong flares can heat dust, pushing the sublimation zone outward. The strongest flare activity is observed on M dwarfs, with peak luminosities that are a few times that of the total quiescent starlight in rare cases \citep[``hyperflares'', see][]{chang2018}. For earlier-type stars, the increase in the bolometric luminosity during ``superflares'' is a fraction of the quiescent level \citep{shibamaya2013, katsova2018}. (Flare amplitudes can be much greater relative to quiescent starlight when observed in specific wavebands.) The radial extent of the sublimation zone during a flare would increase by $(1+A)^{1/2}$ where $A$ is the peak amplitude of a flare in units of the quiescent stellar flux. If dust grains are destroyed by strong flares faster than they can be replenished by other mechanisms (e.g., radial drift from Poynting-Roberston drag), then flare activity increases the radius of the the sublimation zone by as much as a factor of roughly two for the most active red dwarfs.

Here, our goal is to seek evidence of dusty debris throughout a planetary system, wherever it resides at the time of observation, not just near the sublimation zone. Toward that end, we proceed to a parameterized description of the distribution of dust within debris disks.

\subsection{Dusty disk models}\label{subsec:disk}

The next step toward quantifying reflected light echoes from a dusty circumstellar disk is to model the mass distribution within it. We adopt a simple set of disk parameters that include an inner radius $\ain$, an outer radius $\aout$, along with a surface density (mass per unit area), and scale height, respectively:
\begin{eqnarray}
\label{eq:sigma}
    \Sigma(a) & = &  \Sin \left(\frac{a}{\ain}\right)^{\!-\gamma},
    \text{\ \ \ and} \\  \label{eq:h}
        H(a) & = & h \, \ain \left(\frac{a}{\ain}\right)^{\!\beta}, 
\end{eqnarray}
where $a$ is a radial coordinate, $\Sin$ is the surface density at the inner edge of the disk, and $h$ is an aspect ratio defined so that the disk's vertical scale height above the midplane is $h\ain$ at the inner edge of the disk. Here, we focus on geometrically slim disks, with $h \sim 0.05$. The power-law indices are $\gamma \approx 1$--2, and $\beta \approx 1$--1.25, where larger values correspond to a flared disk \citep[e.g.,][]{kenyon1987}. 

With the choices $\gamma = \beta = 1$, total disk mass is
\begin{equation}\label{eq:mdisk}
    \mdisk = 2\pi \int_\ain^\aout \Sigma(a) a da
    = 2\pi \Sin \ain \left(\aout-\ain\right).
\end{equation}
Thus, there is an equal amount of mass in annuli of equal radial width, and the reflecting area of the dust in those annuli is identical also. 

It will turn out to be convenient to relate the surface density of the disk to the global parameters $\ain$, $\aout$ and $\mdisk$:
\begin{eqnarray}
\Sigma(a)   & = &   \frac{\mdisk}{2\pi a (\aout-\ain)}, \\
\label{eq:numperarea}
\frac{\Sigma(a)}{\massp} & = & \mdisk \frac{3}{40 \pi^2 \rhop \rratfac \rmin^3 a (\aout-\ain)},
\end{eqnarray}
where the lower expression is the number of disk particles per unit area projected to the disk midplane.

%


\subsection{Total reflected starlight}\label{subsec:reflect}

With the results in the preceding subsections, we have the ingredients to estimate the light echo from a dusty disk. Our calculations, in (\S\ref{sec:echoes}) below, focus on light curves from systems where the geometry of the disk and its orientation relative to a distant observer affect the arrival time of light reflected from various regions of the disk. Before proceeding to those calculations, we consider the total amount of starlight intercepted by a disk to estimate the magnitude of potential light echoes. From results in \S\S\ref{subsec:dust}--\ref{subsec:disk}, 
\begin{eqnarray}
    f_{\text{total}} & = & \int_{\ain}^{\aout} \frac{2\pi\Sigma(a)}{\massp}\frac{\Ap}{4\pi a^2} da 
    \ \ \ \text{(optically thin disk)} 
\\ & \approx & 
\frac{3\mdisk}{16\pi\rhop\rratfac\rmin\ain\aout} \\ 
\label{eq:ftot}
    f_{\text{total}} & \approx &  
    0.014 \left[\frac{M}{0.1~M_\oplus}\right]
    \left[\frac{\rhop}{2~\text{g/cm}^3}\right]^{-1}
    \left[\frac{\rmin}{1~\text{\micron}}\right]^{\!-1/2}
    \left[\frac{\rmax}{1~\text{mm}}\right]^{\!-1/2}
    \left[\frac{\ain}{30~\text{au}}\right]^{\!-1}
    \left[\frac{\aout}{60~\text{au}}\right]^{\!-1}
\end{eqnarray}
where we have assumed that the disk is optically thin --- starlight is not attenuated as it travels radially through the disk. The middle equation casts the reflected light in terms of the total mass of the disk as well as its overall geometry. The bottom equation provides a numerical value for order-of-magnitude estimates of dust and disk parameters for comparison with observations \citep[e.g.,][and references therein]{hughes2018}. 

In the limit of an optically thick disk that is geometrically thick as well, the total amount of reflected starlight is determined by the height of the disk \citep{gaidos1994}: 
\begin{eqnarray}
    f_{\text{total}} & = & \frac{2\pi a \times 2 H}{4\pi a^2} = h
\ \ \ \text{(optically thick disk)},
\end{eqnarray}
where we have assumed that the disk's height at its inner edge is much larger than the radius of the central star.
Reflected light then comes exclusively from the inner rim of the disk.  An infinitely extended, optically thick disk that instead is geometrically thin intercepts 25\% of the stellar light \citep{adams1986}, while flared disks may intercept as much as 50\% \citep{kenyon1987}.

Generally, scattered light will be brighter near the inner edge of the disk where the number density of particles is higher. When the optical depth is small compared with the inner radius, only the inner edge of the disk scatters light \citep{gaidos1994}. As a rule of thumb, if estimates of the fraction of reflected light in the optically thin limit (Eq.~(\ref{eq:ftot})) are comparable to the disk scale height parameter $h$, then the disk is optically thick, requiring a more detailed analysis (see Appendix~\ref{appx:tau}). From Equation~(\ref{eq:mthresh}), there is a threshold disk mass, depending in disk geometry and dust properties, that marks the transition between optically thick and thin disks: 
\begin{eqnarray}\label{eq:mthresh}
M_\tau & \sim & \frac{16\pi}{3} \ain \aout h \rhop \rratfac \rmin \\
       & \approx & 0.36 \left[\frac{\ain}{30~\text{au}}\right]
       \left[\frac{\aout}{60~\text{au}}\right]
       \left[\frac{h}{0.05}\right]
       \left[\frac{\rhop}{2~\text{g/cm}^3}\right]
       \left[\frac{\rratfac}{31.6}\right]
       \left[\frac{\rmin}{1~\micron}\right] \ \Mearth
\end{eqnarray}
(see Eq.~(\ref{eq:appxmthresh})). For the analysis in \S\ref{sec:kep}, below, focusing on exosolar analogs of the Kuiper belt, $\aoutmax \sim 90$~au; for a ring at that radial distance, a mass of $\sim 2$~\Mearth delineates optically thick and thin disks.  The corresponding mass limit for optically thin disks in the terrestrial zone ($\aoutmax \sim 4$~au), also considered in \S\ref{sec:kep}, is roughly 0.02~\Mearth. Optically thin disks with greater masses are possible if their scale height parameter is increased from the fiducial value of $h = 0.05$.

To provide examples of light reflected from observed circumstellar dust, we adopt parameters loosely based on the well-known debris disk observed around the nearby M-dwarf AU~Mic \citep{kalas2004}. We choose a disk with  0.01~M$_\oplus$ in dust with sizes in the range of 0.1~\micron\ to 1~mm, a radial extent of $\ain = 20$~au to $\aout = 40$~au, a shallow surface density fall-off ($\gamma=1$), and a linear rise in disk scale height with orbital distance ($h = 0.05$, $\beta=1$); \citep[e.g.,][]{strubbe2006, matthews2015, daley2019}. The total fraction of light intercepted by dust is formally about 1\%\ of the starlight in the optically thin approximation\footnote{If the scale-height factor $h$ were reduced significantly, $h \lesssim 0.02$, it would be optically thick in the radial direction. Then we would apply a correction for the attenuation of light propagating through the disk, as described in \S\ref{appx:tau}}. 

For a second illustration, we consider the young, 1.6~\Msolar\ star HD~95086 \citep{su2015}, with a cold 0.18~\Mearth\ debris disk of grains up to a millimeter in size that extends from 60--190~AU. (The total mass of the disk may be as high as 100~\Mearth). The reflected light from this disk potentially amounts to 0.4\%\ of the starlight.

These examples provide estimates of the total reflected light averaged over all observer directions and integrated over time. The reflected light echo from a flare as measured by a distant observer depends on the inclination of the disk and how dust grains scatter photons (Eq.~(\ref{eq:phasefn})). Furthermore, the varied arrival times of echoes from different regions of the disk yield a distinct profile in the observed post-flare light curve. Next, we consider these effects toward distinguishing reflected photons from starlight in the time domain.

\section{Echo light curves}\label{sec:echoes}

\citet{gaidos1994} led the way with predictions of light echoes from optically thick, flared disks based on analytical estimates. For a $\delta$-function flare, the light curve of the echo is
\begin{equation}\label{eq:gaidos}
    f(t) \sim \left(\frac{t}{\tau}\right)^{-2}
    \times \begin{cases}
    1 & \text{if } t\ge \tau \\ 0 & \text{otherwise}
     \end{cases},
\end{equation}
where $\tau$ is the light-crossing time from the star to the inner disk edge. While the expression holds only for a limited set of conditions, including a face-on orientation relative to the observer, this expression serves as a guide to how light echoes are spread out in time. 

\subsection{The echo of a single flare}

Here we modify the estimate in Equation~(\ref{eq:gaidos}) using the results in \S\ref{sec:bg} and a numerical method based on a two-dimensional grid in the plane of the disk. We choose the indices $j$ and $k$ to refer to radial and azimuthal grid points. We assume the disk's scale height is small compared with its radial extent and do not include a third index, corresponding to a polar coordinate, which could be included for a geometrically thick disk or torus. We construct a disk using volume elements $\dV_{jk}$, located at position $\vec{a}_{jk}$ relative to the host star, and containing some amount of dust as specified in a disk model. In general, the delay time for light reflected off each element to reach the observer is
\begin{equation}\label{eq:echodelay}
    \tau =\frac{1}{c} \left(|\vec{d}| -  \vec{d}\cdot\unitobs\right),
\end{equation}
where $\vec{d}$ is the position of the volume element relative to the locus of the flare, $\unitobs$ is the unit vector pointing to the observer, and $c$ is the speed of light. By assuming that the flare locus is at the origin (i.e., that the disk's radial size is much larger than the stellar radius), we can map this expression onto the grid to get the time delay for the $jk$-th element:
\begin{equation}
    t_{jk} =\frac{1}{c} \left({a}_{j} - \vec{a}_{jk}\cdot\unitobs\right),
\end{equation}
where $a_{j} = |\vec{a}_{jk}|$ (independent of azimuthal index $k$). When the disk is geometrically thin, and its inclination is $\inc = 0^\circ$ (face-on), this expression reduces to $t_{jk} = a_{j}/c$. 

With echo delay times mapped from every volume element, the light profile of a star following a flare is straightforward to estimate. For an impulsive flare at time $t=0$, the light curve of an echo is measured in time bins of duration $\Delta t$ and centered at times $\{t_i=i\Delta t\}$. The flux from an echo that lands in time bin $i$ is 
\begin{equation}\label{eq:echogrid}
\Fechoi \propto 
\sum{j,k} C_{jk} F_{i,jk} = 
\sum_{j,k} C_{jk} \frac{\Sigma(a{j}) dA_{jk}}{\massp} \frac{\Ap}{a_{j}^2} \phasefn(\theta_{jk}) 
\times 
\begin{cases} 1 & \text{if}\ |t_{jk}-t_i| < \Delta t/2 \\
0 & \text{otherwise}\\
\end{cases}
\end{equation}
where $dA_{jk}$ is the surface area of the disk (projected onto its midplane) represented by the ${jk}$-th grid point so that the leftmost fraction in the summation is the number of disk particles there, and $\theta_{jk} = \hat{a}_{jk}\cdot\unitobs$ is the scattering angle of light observed from that part of the disk. The factor $C_{jk}$ accounts for any attenuation of starlight by elements of the disk closer to the star as well as the visibility of a flare from the $jk$-th volume element. In the case of an isotropic flare illuminating an optically thin disk, $C_{jk} = 1$ for all $j$ and $k$. However, if flares are localized, originating at random locations on the host star's surface, they illuminate only half of the disk per event. Indeed, only half of a star's flares in this picture are detected at all by a remote observer. In these cases, 
\begin{equation}
    C_{jk} =  
    \begin{cases} 1 \ \ \ (\vec{r}_f \cdot \vec{a}_{jk} \geq 0)
    \\ 0 \ \ \ (\text{otherwise})
    \end{cases}
\end{equation}
where $\vec{r}_f$ is the (random) location of a point-like flare on the stellar surface relative to the star's center. 

\begin{figure}[tbp]
    \centering
    \includegraphics{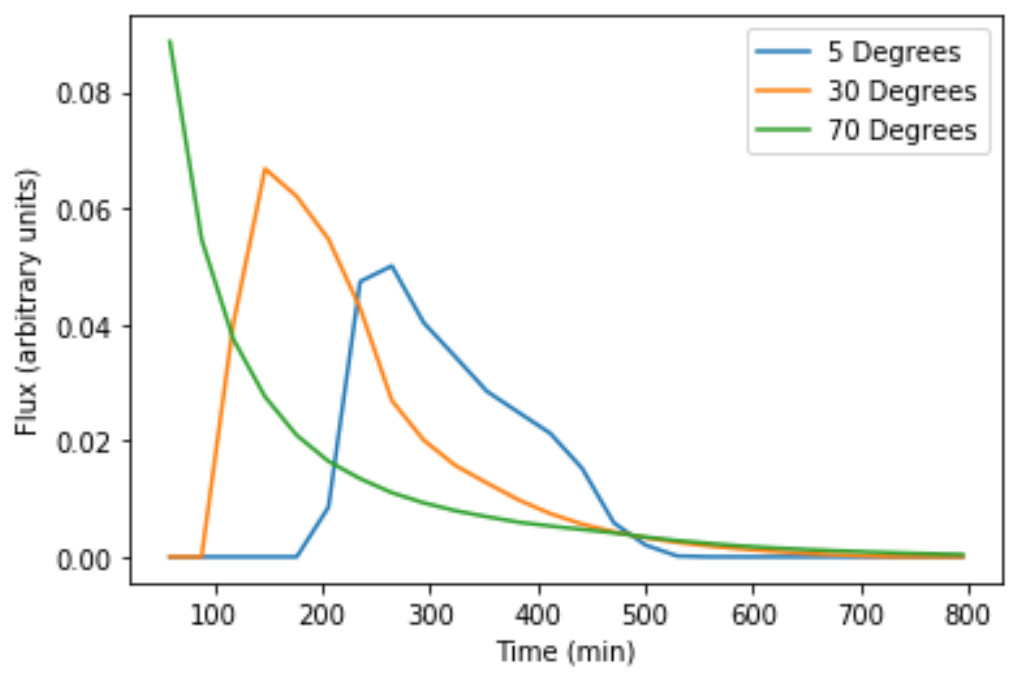}
    \caption{Simulated echo light curves produced by an extended disk, showing the effect of the inclination of the disk relative to the observer. The light curves were calculated for a disk of inner radius 30 AU, outer radius 60 AU, mass of 0.5 Earth masses, and g and $\nu$ factors for the Drain function of 0.429 and 0.114, respectively \citep[see][]{draine2003}}
    \label{fig:echoes_vs_inc}.
\end{figure}

\begin{figure}[tbp]
    \centering
    \includegraphics[width=0.48\textwidth]{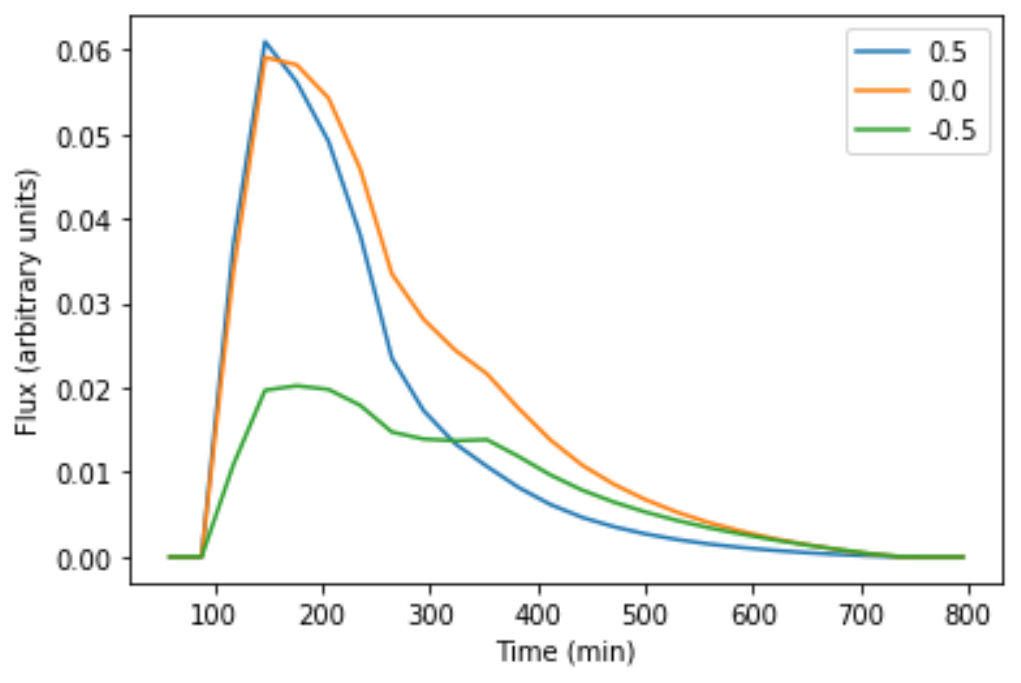} 
    \caption{Echo light curves from a dusty disk showing the dependence on the scattering phase function. The curves illustrate forward scattering ($g = 0.5$), back-scattering ($g = -0.5$), and isotropic scattering ($g = 0$) for a disk with inner/outer radius of 30/60 AU and mass of 0.5 Earth masses, inclined at $30^{\circ}$. Here, the \citet{draine2003} phase function is adopted with $\nu = 0.114$.}
    \label{fig:g_values}
\end{figure}

Figures \ref{fig:echoes_vs_inc} and \ref{fig:g_values} provide illustrations of the echoes from impulsive flares. The first shows the effect of the viewing angle of the disk. Generally, the echo light curve of an inclined disk is broader than when the disk is viewed face on ($\inc = 0^\circ$). Echo profiles are also impacted by the physics of light scattering; forward scattering up-weights echoes from dust in the part of the disk that is nearest to the observer and arrives soon after the flare. Back-scattered light is observed from material on the far side of the disk at later times.

Figures \ref{fig:echoes_vs_inc} and \ref{fig:g_values} suggest guidelines for echo detection from dusty disks. The echo signature may have a distinct peak, but it will generally be broad as a result of differences in the light travel times from various partss of the disk. Thus, if the cadence of an observed light curve is minutes or less, echoes from disks within 1~au of a host star are resolvable. Longer integration times, as in Kepler's 30-minute``long-cadence'' light curves, can resolve echoes from dusty disks that are larger than $\sim 10$~au. Thus the Kepler long-cadence observations can probe time scales that are commensurate with echoes from the exoplanetary analogs of the Sun's Kuiper belt. The one-minute sampling of Kepler's short-cadence data allows us to consider dust in the terrestrial zone of exoplanetary systems.

\subsection{Averaging echoes of multiple flare events}

Flare echoes are likely weak contributors to light curves following a flare. Echoes may be easily lost in low-level stellar activity or instrumental noise. A strategy to enhance the contribution from echoes relative to the noise is to combine light curves from multiple flares, aligning the flare peaks and averaging flux in subsequent time bins.  In a composite light curve built from many flare events, the echo contributions add together, whereas random contributions from noise will wash out. 

In the scenario where flares randomly illuminate half of an axisymmetric disk with every event, averaging dozens or more echo light curves gives an average profile of 
\begin{equation}\label{eq:fmulti}
   \Fmultii \equiv \left< \Fecho(t_i) \right> \approx
   \sum_{jk} (1-\phi_{jk}/\pi) \times F_{i,jk}
\end{equation}
where $F_{i,jk}$ is the echo contribution to the flux in time bin $t_i$ from the region in the disk at some location $\vec{a}_{jk}$  (see Eq.~(\ref{eq:echogrid})) and $\phi_{jk}$ is the azimuthal coordinate of that region, defined so that points in the disk at $\phi = 0$ are nearest to the observer. This weighting assumes that the radius of the star is small compared to the disk dimensions, and that flare events that take place on the far side of the star from the observer are not included in the average.

\subsection{Echo detection strategies}\label{subsec:meth}

The light curve of a star, including an echo from circumstellar material, can be written in the following idealized form:
\begin{equation}
\Flux(t) = Q(t) + \Fflare(t-\tflare) + (\Fflare*G)(t-\tflare) + N(t),
\end{equation}
where $Q$ is the quiescent stellar flux including background variability, $\Fflare$ is the flux from a flare that peaks at time $t=\tflare$, $G$ is an echo structure function that includes the time delay such that $\Fecho=(\Fflare*G)$, and $N(t)$ represents noise, presumably dominated by photon counting statistics. In practice, the light curve is resolved into a series of time-integrated fluxes, with each bin representing the light integrated over some time bin of duration $\Delta t$ plus some additional detector read noise, and minus some additional flux lost between readouts.  Some pre-processing of raw light curves is necessary to remove instrument effects or drift in the detector sensitivity, for example. Often, some forms of processing are accounted for in the data products, such as the PDCSAP flux in the Kepler database; in this analysis, we assume that step has already occurred.  To process light curves of this form, we take the following additional steps:

\begin{itemize}
\item

\textbf{Flatten the light curve.}
We first remove effects like stellar rotation by using a high-pass filter with iterative masking.  Because large impulses like flares can cause ringing effects when processed by digital filters, we include a sigma-threshold mask to hide all large features from the filter, then high-pass the data.  The sigma-mask is then made again and the filter process is applied again to the raw data with the new mask.  Typically three iterations are performed.  Any star with significant residual background variability after this step is likely too variable to be used for echo detection, so we also perform separate screening steps (\S\ref{subsec:sample}).  The resulting light curve is normalized by the filtered background such that the new quiescent flux is unity.  We note that the selection of the filter should be weighed against anticipated echo shapes: because the echo will be longer than the flare, an overly aggressive high-pass filter can remove echo signatures. {See Appendix~\ref{appx:sensitivity} for additional discussion of filtering.}

\item \textbf{Seek clean, impulsive, unresolved flares.}
We then build a flare catalog.  For simplicity, we focused on `delta-flares': events that are not time-resolved, and appear as an excess in a single time bin.\footnote{We have also evaluated processing strategies for resolved flares, but for the case of the Kepler database we found that delta-like flares are far more common and did not pursue them; a faster cadence, however, would likely require such analysis. Some of these considerations are discussed in Appendix~\ref{appx:unresolved}.} These flares, and several time bins before and after the peak, are cropped from the full light curve.  Thus, flares are just single blips in a time series, mathematically represented as Kronecker delta functions, $\Fflare \delta_{ij}$; the flare has its peak magnitude if bin indices $i$ and $j$ are equal, and is zero otherwise. For Kepler long-cadence data, this condition means we seek strong flares that last minutes, not hours. We select these types of flares by requiring that the flux in time bins adjacent to a flare event are statistically consistent with quiescent starlight. 
There are disadvantages to this approach: an unresolved flare is a flare that is thoroughly mixed with background signal, reducing the signal-to-background ratio.  Furthermore, blips from high energy particles can appear as a single-bin amplitude excursion, though these events can often be detected from a point-spread function analysis.

\item \textbf{Scale individual flares to a peak amplitude of unity.}
The echo is proportional to the flare magnitude (and a phase function), so strong and weak flares should, on average, produce proportionally equal echoes. Therefore, we subtract the background and scale the cropped flare catalog so that every flare event has an integrated flux of unity. Now, each cropped light curve has a flare with a peak flux of unity and the rest of the values should be centered about zero if there is no additional activity and no echoes; for flare occurring in bin $k$,
\begin{equation}
    \Fluxnorm = (\Flux - Q)/(\Flux_k - Q),
\end{equation}
where $\Flux_k$ is the flux in the time bin at the time of the flare event. With the additional change of units such that the duration of the time bins is $\Delta t = 1$, the integrated flux of the flare above the stellar background is also unity. 
Thus, as a cropped light curve, $\Fluxnorm_k=1$ and $\Fluxnorm_{k+i}\sim0$ for $i>0$ up to the next flare in the light curve.

\item \textbf{Generate a composite light curve by averaging multiple light curves together.} 
We generate an average light curve for a star, built from individual profiles aligned so that flare events are all at time $t_0 = 0$, and fluxes are averaged within the subsequent time bins $\{t_i\}$. Uncertainties in the mean fluxes in these bins can be estimated using resample techniques with outlier rejection (see \S\ref{sec:kep}). While a straightforward average is possible because the flares are all normalized, their signal-to-noise ratios are not the same.  Large flares will have higher-quality echoes, while smaller flares that barely rise above the noise threshold themselves will have the same peak normalized value of 1, effectively amplifying the noise in the post-flare indices.  We compensate for this by performing a weighted average.  There are different ways to produce weights; one simple weight is the background-relative flare magnitude divided by the standard deviation of the flare-normalized post-flare values.

\end{itemize}

This last step is critical to a search for echo detection as echoes will be weak and may be difficult to distinguish from noise or stellar activity following a strong flare. The echo of an impulsive, unresolved flare from a dusty disk is typically distributed over multiple time bins. Even when dust can reflect 5\%\ of the starlight, the peak flux from the echo may be less than 1\%\ of the flare's contribution because of different light travel times (e.g., Fig.~\ref{fig:echoes_vs_inc}). Thus, an echo detection strategy should focus on broadly distributed excess flux above the quiescent background, not just individual peaks in the light curve following a flare. Outlier rejection and cross-validation methods can mitigate the more severe impact of low-level, micro-flares, 
while random noise will cancel out and weaken as faint echoes build up coherently from many flare light curves combined together. 

The output of this procedure is the composite post-flare light curve of an individual star. To seek evidence of echoes, we fit this composite light curve with echo profile models as in Equations~(\ref{eq:echogrid}) and (\ref{eq:fmulti}), taking into account the uncertainties in the rescaled, averaged fluxes. By mapping the goodness of fit to data in the parameter space of the models, we determine statistically acceptable parameter values. We focus on the total disk mass, a measure of echo strength, to assess if disk echoes are present, and at what level.  Here are details: 

\begin{itemize}
    \item \textbf{Choose fitting parameters.}
Our fitting parameters correspond to disk mass, geometry and orientation. We assume a shallow surface density fall-off ($\gamma = 1$), no flaring in the disk's vertical extent ($\beta = 1$), and a fixed set of  dust properties (e.g., grain size distribution with $q=3.5$ and phase function as in Eq.~(\ref{eq:phasefn})). The remaining parameters for fitting are ($\mfit,\ain,\aout,\inc$), where the first one is the disk mass,
\begin{equation}\label{eq:mfit}
    \mfit = \mdisk \left[\frac{\rratfac}{31.6}\right]^{\!-1}\left[\frac{\rmin}{1~\micron}\right]^{\!-1}
    \left[\frac{\rhop}{2~\text{g/cm}^3}\right]^{\!-1},
\end{equation}
and we have set the fiducial value for $\rratfac$ to correspond to a particle distribution with $\rmax/\rmin = 1000$. With $\rmin = 1$~\micron\,  this choice means that $\mfit$ gives the mass of millimeter-sized grains, which can be compared with observations based on scattered light or reprocessed, thermal dust emission \citep[e.g.,][]{su2015}. 

\item
\textbf{Define a likelihood function.}
For any choice of these model parameters, a measure of goodness-of-fit is 
\begin{equation}\label{eq:chi2}
\chi^2 = \sum_{i=1}^{N} \left[\Fluxnorm_i  - \Fmulti(t_i - t_0, \mfit, \ain,\aout,\inc)\right]^2/\sigma_i^2
\end{equation}
where $i$ is the index for time bins, and  $t_0$ is the time associated with the bin containing the flare event. With the assumption of independent Gaussian errors for the observed flux in time bins, the logarithm of the likelihood function for each model realization is
\begin{equation}\label{eq:likelihood}
    \ln {\cal P} = -\frac{1}{2} \chi^2 + \ln\left(\sum_i 2\pi \sigma_i^2\right) + \ln {\cal P}_\text{priors}
\end{equation}
where ${\cal P}_\text{priors}$ is the prior function, containing our assumptions about the parameter values:
\begin{equation}
    {\cal P}_\text{priors} \sim  
    \begin{cases}
    \ \sin(\inc) \exp(-\mfit/M_\tau) & \ \ 0 \leq \inc \leq \pi/2,  
    \ 0 \leq \mfit/\Mearth, 
    \ \ \text{and} \\
    \ & \ \ \ \ \ \ \ainmin \leq \ain < \aout \leq \aoutmax, 
    \\
    \ 0 & \ \ \text{otherwise.}
    \end{cases} 
\end{equation}
This function encodes our assumptions that the orientation of disks is random, while $\ainmin$ and $\aoutmax$ are limits on the inner disk radius set by the temporal resolution and the duration of the post-flare light curves. Here, we choose $\ainmin = 2 c \Delta t$ to avoid preprocessing artifacts (e.g., detrending). To ensure that all model light curves otherwise produce echoes that do not exceed the maximum observed delay time, $T_\text{max}$, the outer radius of the models are limited to $\aoutmax = c T_\text{max} / 2$. The priors also include an exponential cut that limits disk masses that correspond to optically thin disks.

\item
\textbf{Map the likelihood in parameter space.}
We use the Markov-chain Monte Carlo (MCMC) routine \mysoft{emcee} \citep{foreman-mackey2013} to generate samples of the likelihood distribution in Equation~(\ref{eq:likelihood}). Python's \texttt{minimize()} routine in the \texttt{scipy.stats} module finds parameters that minimize $\chi^2$, which help to seed the MCMC sampler. These samples enable estimation of expectation values and confidence regions in parameter space. 

\end{itemize}

\begin{figure}[tbp]
    \centering
    \centerline{\mbox{\includegraphics[width=0.5\textwidth]{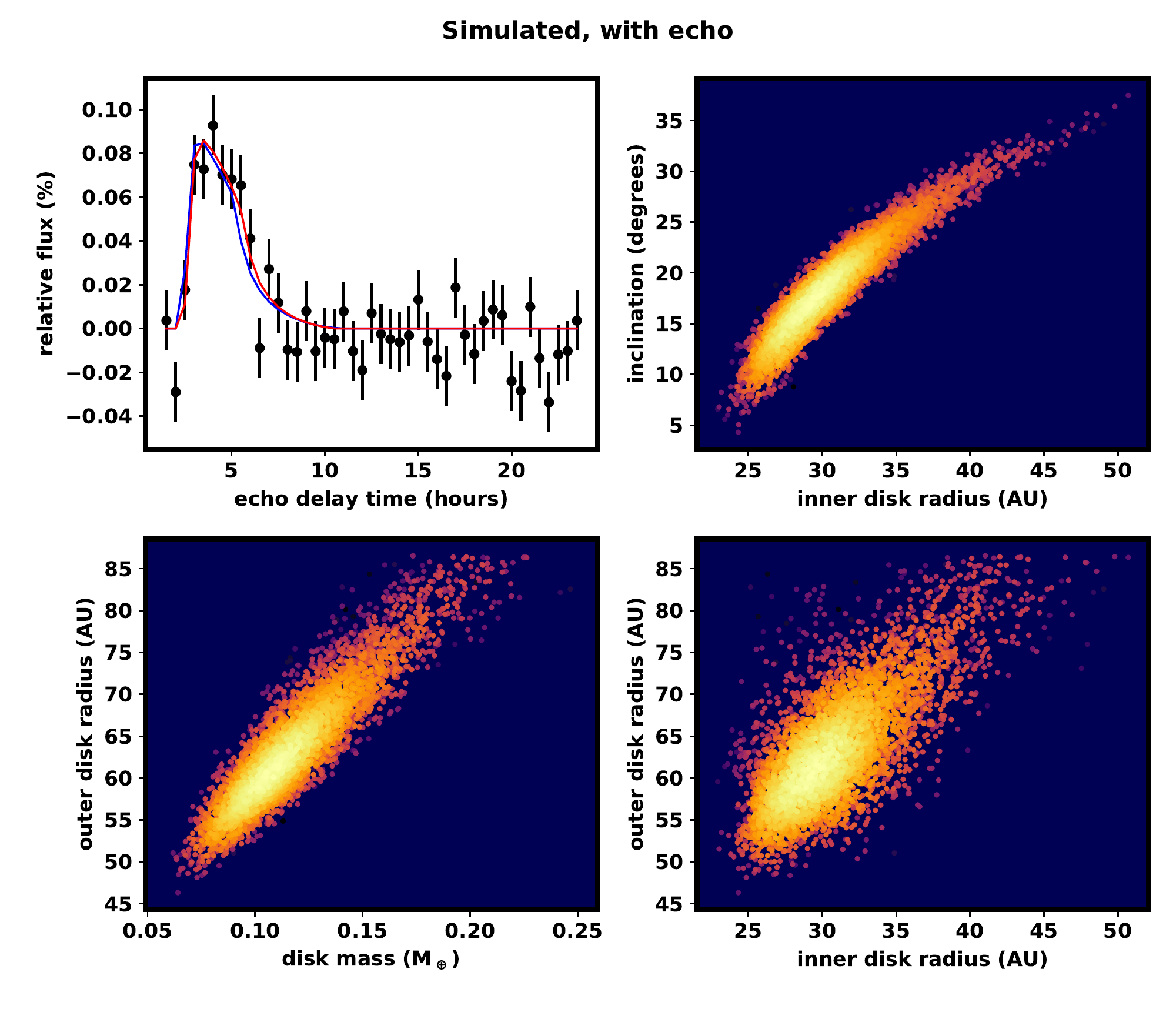}
    \includegraphics[width=0.5\textwidth]{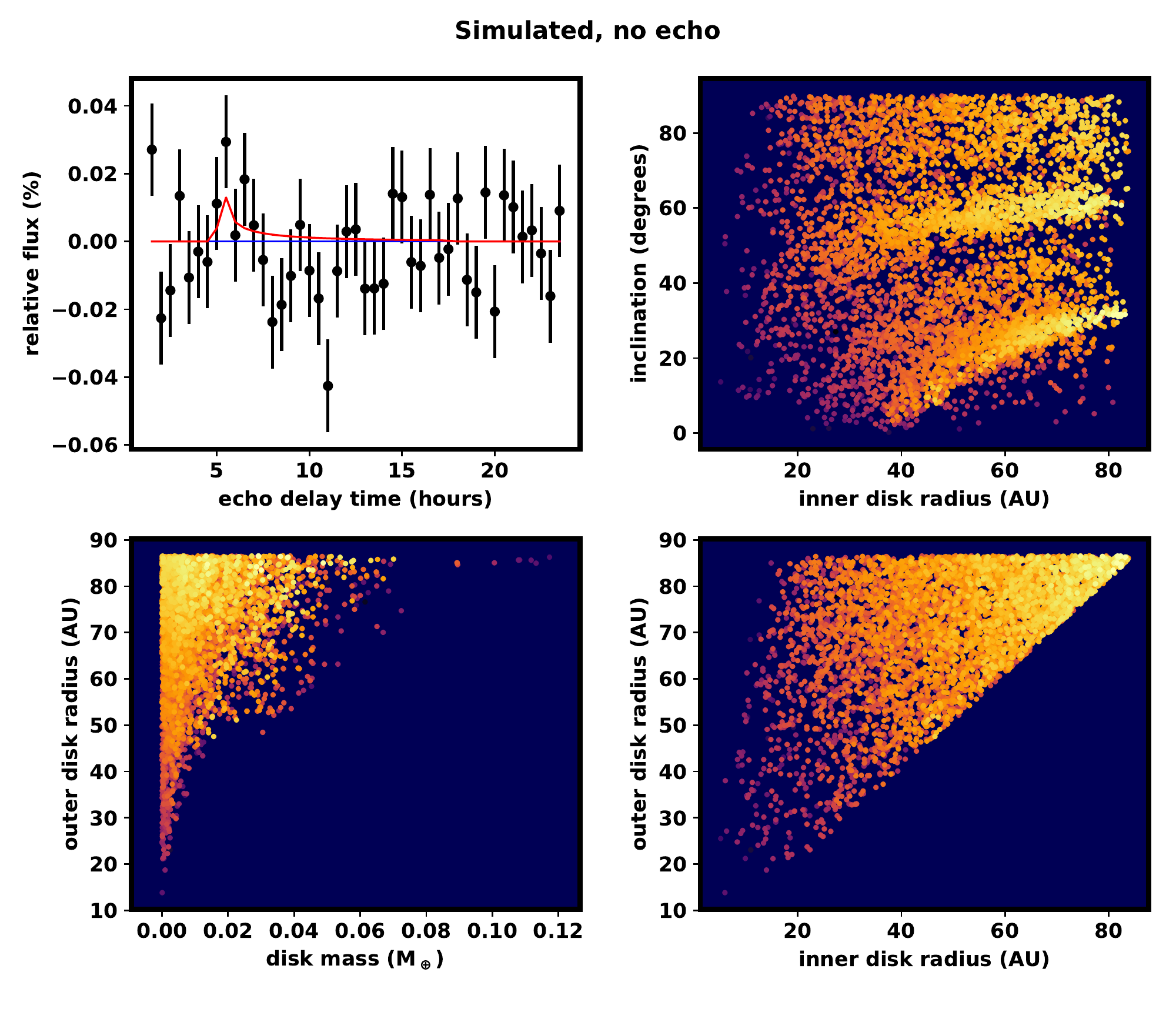}}
    }
    \caption{\label{fig:echofake} 
    Simulated noisy light curves with and without flare echoes. The panels on the left are derived from a model with disk parameters $(\mfit, \ain,\aout,\inc)=(0.1~\Mearth, 30~\text{au}, 60~\text{au}, 20^\circ)$. The upper left is the composite light curve, including the synthetic data (points with 1-$\sigma$ error bars), the true echo (blue curve), and the best-fit model (red curve). The three density plots show the projection of MCMC-generated samples of the likelihood distribution of model fits. The colors correlate with likelihood; brighter colors indicate higher values. The plots on the right correspond to a light curve with no echo. Application of the numerical machinery of MCMC sampling yields disk model parameters that are consistent with the data. While the samples are just ``fits to the noise'' they show that upper limits may be placed on key parameters such as the disk mass.}
\end{figure}

Figure \ref{fig:echofake} provides an example of this echo detection method applied to synthetic data. The left panels show a composite light curve of 30 simulated flare events and echoes from a disk with parameters $(\mfit, \ain,\aout,\inc)=(0.1~\Mearth, 30~\text{au}, 60~\text{au}, 20^\circ)$ compared to a control with no echo on the right.  The figure also provides maps in parameter space showing projections of the likelihood distribution. In the case of a statistically significant echo profile --- here $\sim$0.1\% of the starlight is reflected by the disk --- recovery of the true parameters is feasible. 

A challenge for parameter estimation is that similar echo shapes are predicted for a range of inclination inner disk radii. This approximate degeneracy may be understood in terms of light travel time from the front part of the inner edge of the disk, which, in our forward-scattering dust models is the brightest region. Photon arrival times at a distant observer from the front of the disk are the same in cases where the disk has a small $\ain$ and low $\inc$ as when both $\ain$ and $\inc$ have higher values. Increasing the inclination tilts the front part of the disk toward the observer, compensating for the longer light travel time between the star and a more distant inner edge. This effect leads to the banana-shaped curve in the upper-right panel of Figure~\ref{fig:echofake}, which we note is not present in the no-echo case.

\subsection{Limits on the presence of dust}

A main deliverable of the above procedure is the distribution of echo strengths, quantified by the disk mass parameter $\mfit$ (Eq.~(\ref{eq:mfit})). By definition, $\mfit$ corresponds to a disk of micron- to millimeter-sized dust; different dust sizes are associated with different disk masses.  The disk models adopted here allow for simple transformations between this parameter and other physical quantities like particle sizes, and scattering area of the disk, which is more directly connected with echoes strength. The MCMC samples of the likelihood function in Equation~(\ref{eq:likelihood}) allow constraints on $\mfit$, marginalizing the remaining parameters.  When a light curve contains a statistically significant echo, these samples yield confidence limits for total disk mass in dust. For example, the samples in Figure~\ref{fig:echofake} yield 95\%\ -confidence limits of 0.084--0.172~\Mearth. 

In instances where an echo signal is not statistically significant, model fits provide limits to the mass of dusty disks as allowed by the data; reflected light from larger disk masses would produce fluxes that are inconsistent with the measured light curve.
The right-hand panels of Figure \ref{fig:echofake} show MCMC samples corresponding to fits to a composite simulated light curve that has only noise; ``truth'' is $\mfit = 0$. Disk models with some mass are consistent with the data. The likelihood map in that figure reveals a range of possible disk masses, bounded by $\mlim = 0.026$~\Mearth\ when adopting the condition that 95\%\ of the MCMC samples have masses below this value. We take $\mlim$ to be our formal mass limit for optically thin disks of micron- to millimeter-sized dust with radial extent less than $\aoutmax$. 

Our metric for the presence of dust, a mass limit, is one of many possibilities. The total cross-sectional area of dusty debris is another. As with mass, how this area connects to the shape of an echo profile or the integrated echo strength is dependent on dust properties and disk models. We choose to present results in terms of mass because of the connection with planet formation models and the reservoir of solids available to create debris. The presence of an echo may also be inferred with less reliance on model details. The artifact injection procedure in \S{\ref{subsec:artifact}}, below, is an example that considers mass limits assuming optimal conditions for echo production. Our next step is to use these estimators to hunt for echoes in observed light curves.


\newpage
\section{Analysis of Kepler data}\label{sec:kep}

The exquisite photometric time series of the starlight from over 150,000 stars observed by NASA's Kepler satellite \citep{borucki2010} offers a unique opportunity to search for flare echoes. These data contain the light curves of tens of thousands of active stars \citep{walkowicz2011, hawley2014, yang2017}. Here, we mine the data from stars in the Kepler Input Catalog (KIC) for impulsive flares, seeking the signatures of echoes in composite light curves. 
We begin with the selection of candidate stars.

\subsection{Sample selection}\label{subsec:sample}


\begin{itemize}
    \item 
\textbf{Initial Sample of Stars.}
To identify promising candidate stars, a robust, quarter-by-quarter search of the long-cadence Kepler database was performed. This initial pass was not extremely detailed, and only sought to identify the most active flare stars in the database. Potential candidates were ranked using a weighted average based on two different measures of flare activity, with priority given to highly active stars with powerful flares. Flares were initially counted using a sigma threshold relative to the quiescent flux -- $3\sigma$ for all flares, $6\sigma$ for powerful flares. To help avoid possible false positives due to noise in high-$\sigma$, low-variation stars, a background-percentage measure was also used. Peaks that reached above 1\% of the quiescent flux (based on normalizing light curves by their median flux value) were counted, with 4\% being a measure of powerful flares. On the basis of these measures of flare activity, we identified preliminary sets of 3,000 stars in the long-cadence observation list, and 1,000 stars observed at short cadence.

\item \textbf{Detrending light curves.}
From this initial list of stars, the PDCSAP flux from all available quarters was stitched, normalized, and flattened using \mysoft{lightkurve} (v1.11.2; \citealt{lightkurve2018}).  The flattening procedure wraps \texttt{scipy.signal.savgol{\_}filter}, and was processed with 3 iterations at 3-sigma using 2nd order polynomials with window length 101 (see Appendix \ref{appx:sensitivity}).  This step produces a light curve that is normalized to the quiescent background, with a median of one. The light curves were then screened for residual periodic signal (e.g., from rapid stellar rotation) that could complicate the analysis.  A Lomb-Scargle periodogram was computed with the \mysoft{Lightkurve} \texttt{to{\_}periodogram} function, and any light curves with a maximum power higher than 0.008 were rejected.  Additionally, the autocorrelation of the entire light curve was calculated and any values higher than 0.1 beyond index 8 were cause for rejection.  These steps select for candidate stars that have a slowly varying or constant quiescent state.

\item \textbf{Identifying unresolved flares.}
A preliminary flare catalog was generated by identifying 5- and 9-sigma peaks for the long and short cadence data, respectively. They were further processed by rejecting any flares that had another flare occur within the post-flare analysis window.  The flare peak time was determined by selecting the local \texttt{argmax} index.  
A weight was computed for each flare by taking the peak flare magnitude minus one, divided by the standard deviation of the post-flare analysis window.

Flares were normalized by subtracting one, then dividing by the flare peak value.  This produced a post-flare analysis window nominally centered about 0 with the flare peak having value 1.  A pristine echo peaking at 10{\%} would therefore have a value of 0.1, regardless of the initial flare magnitude or quiescent background.
For the subsequent steps, if the total number of flares remaining in the catalog dropped below 10, the star was dropped from further analysis.  

To simplify analysis, we only included unresolved flares (that is, their structure cannot be determined because their rise and decay occurred in a period that is shorter than the cadence).  Because these flares are fully integrated over, their true peak values are unknown and the fraction of flaring light to quiescent light is lower than with resolved flares.  But they are the simplest case and furthermore, we found that the vast majority of flares in the Kepler catalog can be characterized as delta-like. Of the 3000 top stars in the long cadence data, only 53 stars were characterized as routinely producing resolved flares; 201 stars were characterized as producing both delta-like and resolved flares; 1,784 stars were characterized as predominantly delta-like. The remainder either produced insufficient flares for analysis or their fluctuations later proved to be from a non-flaring origin.

\item
\textbf{Seeking clean, short-duration, unresolved flares.}
To screen for delta-like flares, we rejected any flares that had $>$35{\%} of the flare peak value in $i_{flare}+1$ or $>$12.5{\%} in $i_{flare}+2$.  Outlier rejection included 3 steps—if the flare failed any part of any one step, it was rejected from the catalog.  
\begin{enumerate}
    \item Flares were screened for any values in the post-flare analysis window that had an absolute value exceeding $>$50{\%} of the peak flare intensity; these typically corresponded to a microflare or transit event.  This was a coarse test to help stabilize the subsequent tests.
    \item A leave-1-out (jackknife) outlier analysis was performed, primarily motivated by microflaring events that produce outsized impacts on the analysis, as follows: 
    \begin{enumerate}
        \item Produce a sub-catalog that removes a single flare from the full catalog; 
        \item Compute the weighted mean value for each index in the post-flare analysis window;
        \item Repeat a \& b for all remaining flares;
        \item Compute the standard deviation of the leave-1-out means;
        \item Any flare that causes a 4-sigma deviation in a single bin is rejected as an outlier.
    \end{enumerate}
    \item A leave-2-out outlier analysis was performed, similar to step 2, where each pair of flares is exhaustively tested, this time with a 5-sigma rejection threshold.  This was motivated by pairs of microflaring events that were discovered to conspire to evade the leave-1-out analysis in some flare catalogs.  
\end{enumerate}

The weighted mean was computed for each index of the remaining inlier flares.  A weighted standard error of the mean was computed as $\sqrt{\sum (F_i-\langle F\rangle)^2w^2}$, where the $F_i$ are flare strengths, and weights $w_i$ are re-normalized for each catalog. The resulting weighted means and weighted standard errors were exported for echo analysis.

\end{itemize}

\subsection{The impulse-flare star catalogs}\label{subsec:catalogs}

The result of our screening the long-cadence data is a catalog of 1,734 stars for which composite light curves are calculated. The minimum number of flares per star is 10, and the maximum detected is 148, with a mean of 48 flares per star for a total of 83,056 flare events. As a result of outlier rejection, composite light curves incorporate 64,421 flares, with a mean of 37 flares per star. Additionally, 627 stars with short-cadence data were selected with the same procedure. These stars have a total of 39,881 flare events, with a mean of 63 flares per star. Following outlier rejection, the number of flare events dropped to 34,337, with a mean of 54 flares per star. Table~\ref{tab:stars} lists example stars from both the long-cadence and short-cadence catalogs of stars with unresolved flares, ranked according to the number of inlier flare events that contribute to the composite light curves. The accompanying electronic table contains the complete collection of stars in these data sets. 

\begin{deluxetable}{cccccccccccc}
\tabletypesize{\footnotesize}
\tablecolumns{12}
\tablewidth{0pt}
\tablecaption{Stars with composite light curves.
    \label{tab:stars}}
\tablehead{\colhead{source} & \colhead{cadence} & \colhead{T$_\text{eff}$} & \colhead{$\log$ g} & \colhead{Gaia $d$} & \colhead{Gaia G} & \colhead{B$_P$-R$_P$}  & \colhead{W1-W3} & \colhead{$N_\text{flares}$} & \colhead{$N_\text{inliers}$} & \colhead{$p$-value} & \colhead{$\mlim$}
\vspace{-0.2cm}
\\
\colhead{KIC} &  & \colhead{(K)} & \colhead{(cm/s$^2$)} & \colhead{(pc)} & \colhead{(mag)} & \colhead{(mag)}  & \colhead{(mag)} &  &  & \colhead{(no echo)} & \colhead{(\Mearth)}}
\decimals
\startdata
03525674 & LC & 5237 & 4.568 & 898 & 14.2082 & 1.0911 & 0.637 & 88 & 67 & 0.2387 &  2.816\\
04038250 & LC & 5876 & 4.3 & 1370 & 13.4266 & 0.8972 & 0.001 & 96 & 67 & 0.2371 &  2.075\\
09692084 & LC & 6072 & 4.371 & 1700 & 14.4879 & 0.8596 & -0.185 & 80 & 64 & 0.8317 &  1.596\\
07867214 & LC & 5730 & 3.998 & 786 & 13.3331 & 0.9361 & 0.186 & 74 & 64 & 0.1958 &  2.592\\
06976550 & LC & 5831 & 3.75 & -- & -- & 0.9292 & -0.883 & 73 & 62 & 0.7497 &  2.418\\
05177163 & LC & 5591 & 4.559 & 870 & 14.1525 & 0.991 & -0.077 & 73 & 62 & 0.8823 &  1.877\\
07950066 & LC & 6292 & 4.426 & 1030 & 14.195 & 0.7906 & -0.218 & 78 & 62 & 0.1957 &  2.332\\
10259010 & LC & 6011 & 4.458 & 1560 & 14.1138 & 0.8805 & 0.096 & 76 & 61 & 0.7858 &  1.883\\
11966791 & LC & 5745 & 4.457 & 1300 & 13.8216 & 0.9281 & 0.381 & 66 & 61 & 0.9262 &  1.924\\
09826112 & LC & 6373 & 4.084 & 953 & 13.2789 & 0.7671 & -0.211 & 72 & 61 & 0.0280 &  3.353\\
02444412 & SC & 5650 & 4.5 & 308 & 12.5606 & 0.9021 & -0.664 & 250 & 201 & 0.3094 &  0.0028\\
08554498 & SC & 5937 & 4.012 & 554 & 11.596 & 0.8243 & -0.086 & 239 & 204 & 0.8207 &  0.0016\\
06541920 & SC & 5657 & 4.364 & 659 & 13.7062 & 0.8978 & 0.248 & 238 & 188 & 0.1612 &  0.0037\\
02571238 & SC & 5546 & 4.575 & 220 & 11.8763 & 0.9142 & 0.098 & 224 & 194 & 0.0573 &  0.0024\\
06850504 & SC & 5465 & 4.449 & 285 & 12.4535 & 0.9427 & 0.034 & 219 & 186 & 0.8362 &  0.0027\\
11512246 & SC & 5761 & 4.091 & 856 & 13.4015 & 0.8231 & -0.036 & 210 & 180 & 0.0840 &  0.0030\\
10925104 & SC & 3980 & 4.722 & 261 & 13.7125 & 1.3817 & -0.021 & 209 & 187 & 0.9636 &  0.0016\\
07603200 & SC & 3846 & 4.738 & 67 & 12.4715 & 1.8824 & 0.091 & 209 & 178 & 0.4422 &  0.0019\\
05446285 & SC & 5486 & 4.449 & 381 & 13.1048 & 0.9868 & 0.307 & 206 & 180 & 0.0450 &  0.0035\\
10187017 & SC & 4900 & 4.602 & 108 & 11.4784 & 1.2395 & 0.012 & 196 & 170 & 0.0251 &  0.0023\\
\enddata
\tablecomments{The stellar properties ($T_\text{eff}$ and $\log\ g$) are from \citet{mathur2017}. The Gaia DR2 data include a distance, which is the inverse of the reported parallax, when positive-valued. Columns specific to this work are: the number of flares identified ($N_\text{flare}$); the number used in generating a star's composite light curve ($N_\text{inliear}$; the $p$-value of a light curve being consistent with no echo (constant flux), and the inferred disk mass limit, defined as the 95th-percentile in fitted disk mass from the MCMC samples. This list is drawn from the accompanying on-line, machine-readable table, which includes three additional column: (i) the 5th percentile of the MCMC sample disk masses, and, if known, (ii) the Gaia DR2 source ID, and (iii) the WISE object identifiers. If these entries are tagged with an asterisk, the identification is tentative, based on exactly one source listed in Vizier \citep{vizier2000} in the respective catalog within 2 arcseconds of the KIC source. }
\end{deluxetable}

\begin{figure}
    \centering
    \includegraphics[scale=0.8]{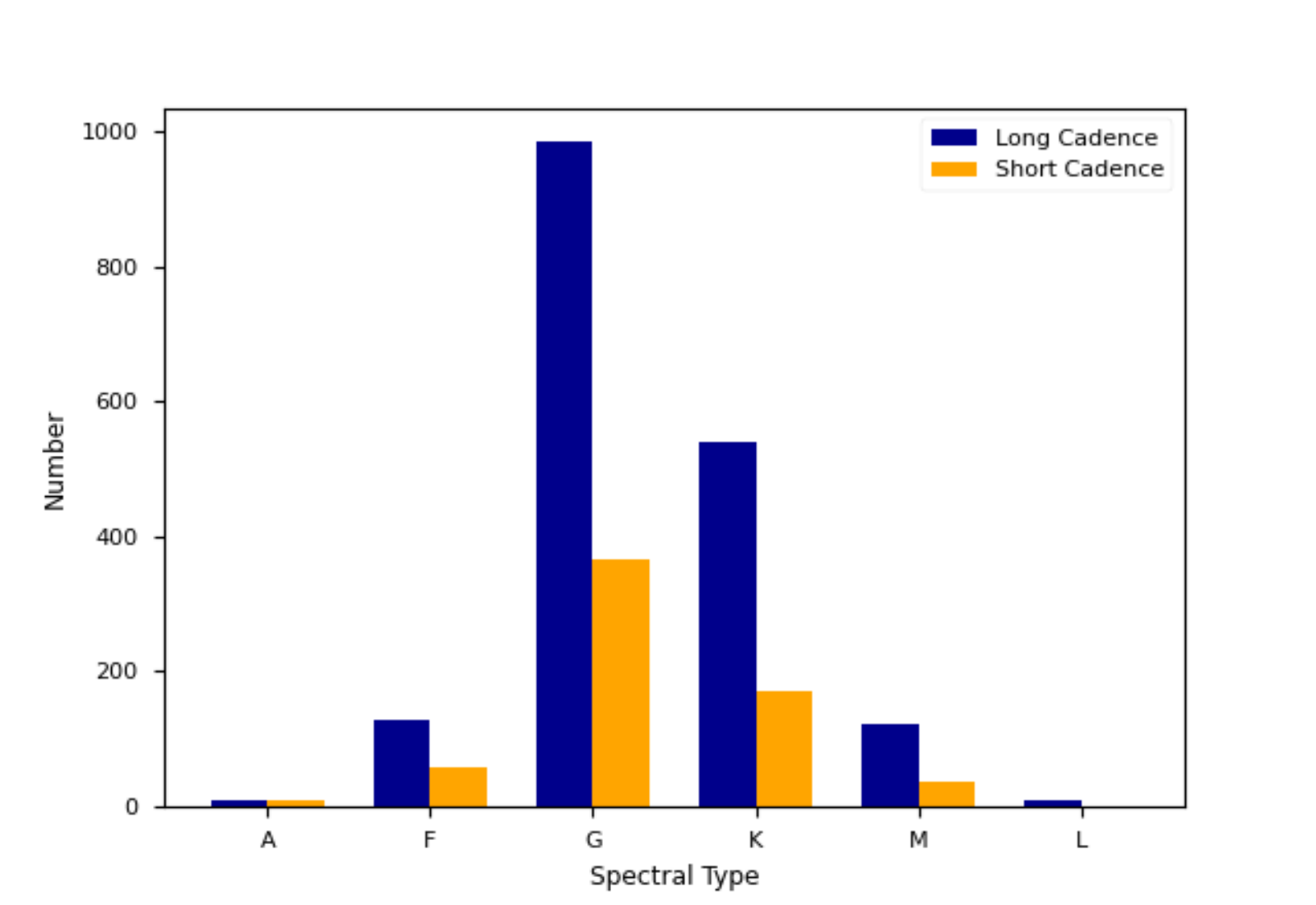}
    \caption{Spectral type distribution of stars in the long-cadence and short-cadence catalogs, based on stellar properties in the Kepler Input Catalog \citep{mathur2017}.
    \label{fig:spectral_type}}
\end{figure}

\begin{figure}
    \centering
    \includegraphics[scale=0.6]{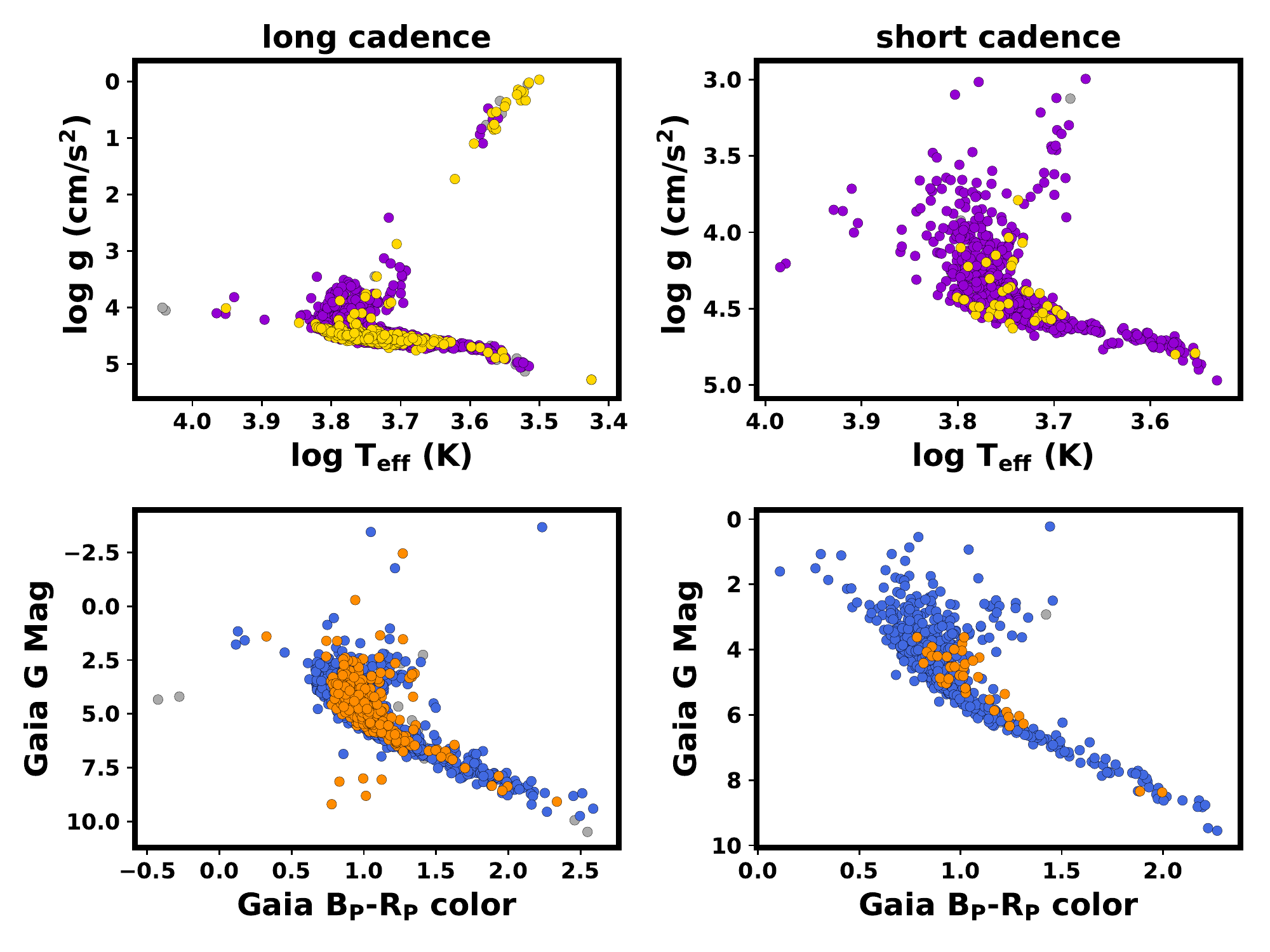}
    \caption{Effective Temperature versus log-surface gravity and color-magnitude diagram (CMD) for the short-cadence and long-cadence catalogs. The upper plots contain stellar parameters from \citet{mathur2017}, while the CMDs on the bottom row are from those objects co-listed in Gaia DR2. The color map is based on WISE infrared color (W1-W3);  stars with a strong IR excess (W1-W3$>$1~mag) appear in yellow and orange, while the purple and blue symbols are for ``bluer'' stars. The color excess may be associated with thermal emission from circumstellar dust  \citep{rizzuto2012}, and thus may help identify candidates for flare-echo studies. We choose different color schemes in the upper and lower panels to highlight that the stellar populations are different. Stellar properties shown in the upper panels are available for most but not all stars in the Kepler Input Catalog, while there are fewer stars with Gaia data in the lower panels. 
    \label{fig:tefflogg}}
\end{figure}

The spectral type distribution of stars in our sample is shown in Figure~\ref{fig:spectral_type}. Stars of spectral type G are most common, comprising over 55\% of both long-cadence and short-cadence data sets. Figure~\ref{fig:tefflogg} has an object-level description; it shows stellar parameters, effective temperature ($T_\text{eff}$) and surface gravity ($\log g$) as determined by \citet{mathur2017}. It also contains color-magnitude diagrams based on distance, photometry and color data from the Gaia DR2 database \citep{gaia2016, gaia2018}. To identify counterparts in Gaia, we used the VizieR \citep{vizier2000} and Simbad \citep{simbad2000} query engines. If a counterpart is known or if a search yields exactly one object within two arcseconds of a Kepler star's sky position in the Gaia DR2 catalog, we associate that source with the KIC star. 

Figure~\ref{fig:tefflogg} also identifies the infrared (IR) color excess  --- a signpost of circumstellar dust --- of the selected active stars.  We identify IR excess using the Wide-field Infrared Survey Explorer (WISE) catalog \citep{wright2010}, focusing on W1-W3 color \citep[cf.][]{rizzuto2012}.  Because the WISE point spread function is large compared to Gaia's, and because of the potential that a WISE object may be extragalactic \citep[see][for example]{kuchner2016}, a WISE source must be within one arcsecond of a KIC star to be associated. Overall, this strategy mitigates confusion in crowded fields, although it may miss some sources.  Identifying sources in the KIC, Gaia DR2, and WISE surveys gives some confidence that we have identified the WISE counterparts of a KIC star. 

In summary, the two catalogs contain data from 2,173 stars, yielding 2,361 composite light curves. For the long-cadence KIC data, our search strategy finds 1,709 Gaia sources, 1,647 WISE sources, and 1,625 stars in both Gaia and WISE. In the short-cadence data, the yield is 625 Gaia sources, 614 WISE objects, and 612 stars with entries in both surveys. 

Based on published observations and stellar properties as in Figure~\ref{fig:tefflogg}, the active stars we select have the following characteristics:

\begin{itemize}
    \item In both the long-cadence and short-cadence catalogs, late-type main sequence stars (F, G, K, M) predominate; these demographics are not surprising given that active stars tend to be late-type stars. Additionally, each catalog has seven A stars; none of these objects are in both catalogs. Some post-main sequence subgiant flare stars are apparent in the color-magnitude diagram as well \citep[e.g.,][]{goodarzi2021}. 
    
    \item In the long-cadence catalog, five Sun-like stars lie below the main sequence. All of these objects are likely main sequence stars with distances that are underestimated in Gaia DR2. In some cases, the parallax errors are large (e.g., KIC~10215638), and in others there is a nearby star that triggers a \texttt{duplicate source} flag (as in KIC~5962506). 
    
    \item The $T_\text{eff}$-$\log g$ diagram distinguishes 28 giant stars ($\log g<2$) that launch powerful flares observed in the long-cadence data \citep{balona2015, vandoorsselaere2017}. The planetary systems they host may offer hints to the origin of planetary material observed around some white dwarfs \citep{farihi2009}.

    \item There are also two hot subdwarfs (KIC~7664467 and KIC~10139564) in the long-cadence data. Flare activity on this class of objects has been reported \citep{balona2015}. The majority of these sources are faint, evolved stars in short-period ($< 10$~d) binaries \citep{maxted2001} with companions that include late-type main-sequence stars \citep{geier2014}. It is unclear whether the lower mass companions generate the flares in these sources.

    \item While the demographics of the long-cadence and short-cadence sources are similar (e.g., Fig.~\ref{fig:tefflogg}, median $T_\text{eff} = 5645, 5718$~K), a greater fraction of long-cadence stars show an IR excess, with 24\% having W1-W3$\geq 1$~mag, compared to 7\%\ of the short-cadence stars. 

\end{itemize}

To conclude, the stars in our long-cadence and short-cadence catalogs are all late-type main sequence stars except for the few dozen objects including a group of evolved red giant branch stars. 

\subsection{A test for the presence of echoes}

As a first approach to interpreting the data from each catalog, we assess whether there is \textit{any} evidence of an echo in a composite light curve. Whether a single spike in flux in one time bin or a sustained flux surplus above the quiescent starlight that spans multiple time bins, a strong echo can be ruled out by a simple test of a ``no-echo hypothesis'' that the flux in a composite light curve is constant in time. We test it by first assuming that fluxes in a light curve have errors that are statistically independent and normally distributed. Using $N_\text{bin}$ flux bins, we compute $\chi^2$, defined as the sum of squared residuals of fluxes in these bins relative to a constant background flux (zero in our rescaled, normalized light curves). We then derive a $p$-value, the likelihood that the observed $\chi^2$ is greater (a poorer fit to a no-echo model) than expected in a theoretical $\chi^2$ distribution with degrees of freedom $dof = N_\text{bin}-1$. (We deduct one because we have already estimated the quiescent flux.) This measure does what we want: the $p$-value for a light curve with a well-defined echo profile is extraordinarily small (it is less than $10^{-20}$ for the data in Fig.\ref{fig:echofake}). When no echo is present, typical $p$-values are of order one-half (0.26 for the data in Fig.~\ref{fig:echofake}). A no-echo hypothesis test amounts to choosing a threshold $p$-value and asking whether measurements lie above that value in support of the hypothesis, or below it, contrary to expectation.

\begin{figure}
    \centering
    \centerline{\includegraphics[width=0.7\textwidth]{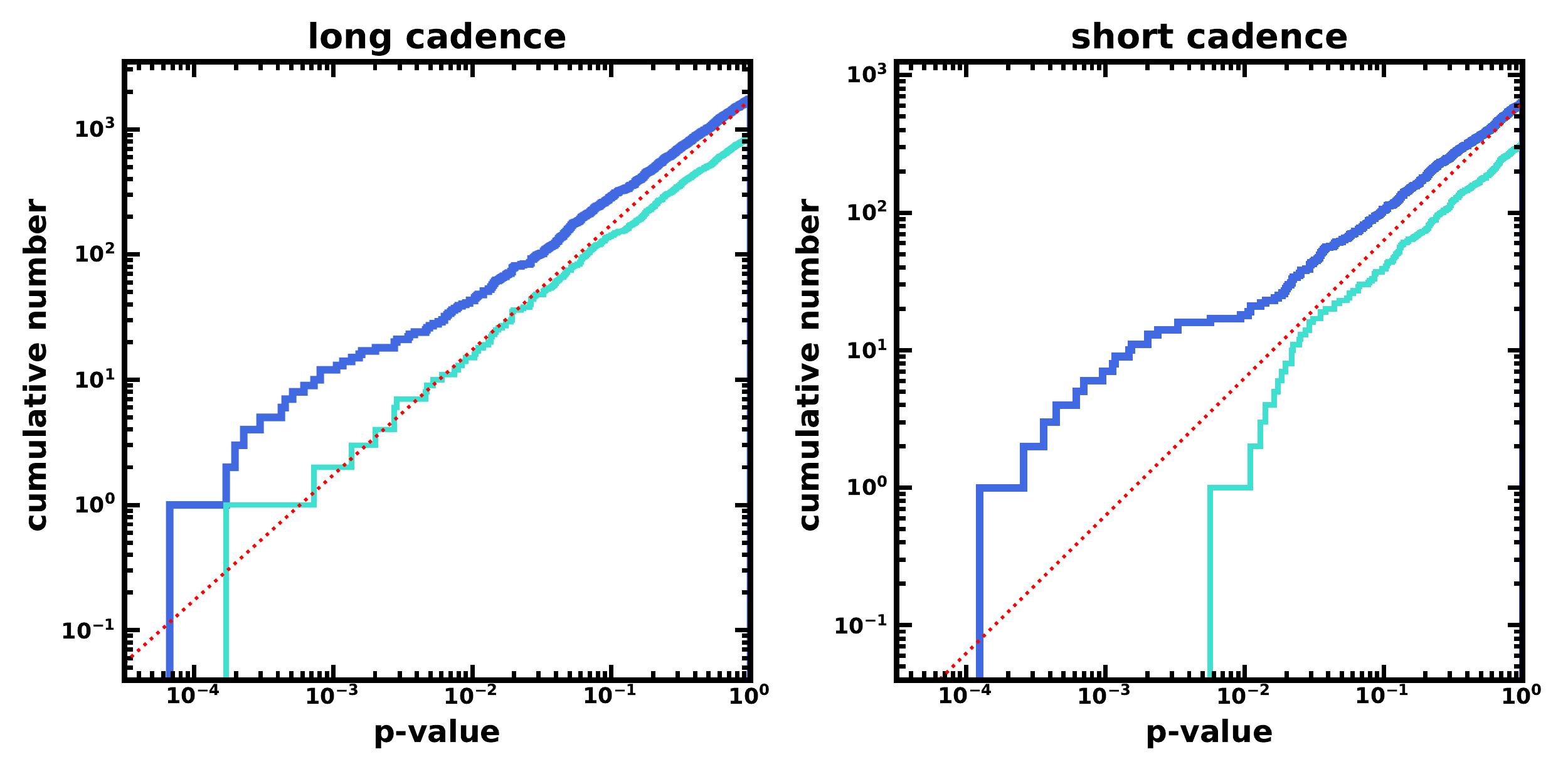}}
    \caption{\label{fig:pvalhist} 
    The cumulative distribution of $p$-values for the no-echo test of long-cadence and short-cadence light curves. The blue histogram are the data; the red dotted line is the expected value, assuming that flux errors are strictly Gaussian and known precisely. The bumps and outliers at low $p$-value relative to the theoretical expectation suggest that the no-echo hypothesis fails more often than expected. We caution that this mismatch is not necessarily either evidence for echoes or a problem with the analysis if there are no detectable echoes. The flux errors in the composite light curves have some dispersion, which shifts the $p$-values relative to an ideal analysis based on known flux uncertainties. To illustrate this point, we sort each data set according to $\nflares$, the number of flare events that contribute to each composite light curve; presumably, the flux uncertainties are better constrained when the number of flares is higher.  We build a second histogram in each panel, shown in cyan, that corresponds to the composite light curves where $\nflares$ is at or above the median value (38 and 44 for the long-cadence and short-cadence data, respectively). The bumps in the full histograms, indicating an excess of low $p$-values, are not present when we consider the better-sampled half of the data. The anomolously low $p$-values --- including the bumps in the histrograms of the full data sets --- are the result of poorly constrained noise, not astrophysics.
}
\end{figure}

Taken as a group, the $p$-values we derive from the composite light curves show the trends we expect when no echoes are present. Figure~\ref{fig:pvalhist} shows the distributions of $p$-values from the long-cadence and short-cadence data sets, which broadly follow the theoretical expectation when there are no detectable echoes.  Still, there is evidence that the no-echo hypothesis fails more often than expected. In both the long-cadence and short-cadence catalogs, there is an excess of composite light curves with low $p$-values, creating a bump in the distribution above the theoretical expectation. There are also individual outliers with unexpectedly low $p$-values, strong rejections of the no-echo hypothesis given the sample sizes. The most extreme case in the short-cadence data has a $p$-value of $1.3\times 10^{-4}$, while the long-cadence catalog has an outlier with a $p$-value of $1.5\times 10^{-6}$. If the assumptions that went into the $p$-value estimate are reasonable (known Gaussian errors in flux bins), then the long-cadence outlier is roughly a one-in-a-million event.

The outliers, including the sources that contribute to the bumps in the distribution of $p$-values for the no-echo hypothesis, may tempt a claim for evidence of echoes. However, by sorting the composite light curves according to $\nflares$, the number of flare events in each composite, we build a case that noise, not astrophysics, is responsible for the excess in low $p$-values relative to the theoretical expectation. Our premise is that a smaller number of flares leads to greater sampling error in the estimate of flux uncertainties, which in turn can strongly affect the derived $p$-values. Indeed, in both the long-cadence and short-cadence data sets, the outliers are entirely associated with composite light curves that have $\nflares$ below the median value (38 and 44, respectively). When only the better-sampled light curves are considered (with $\nflares$ greater than the median), the excess number of low $p$-values --- the bumps in the distribution --- vanish (Fig.~\ref{fig:pvalhist}, cyan histograms). The three most extreme outliers in the long-cadence data are derived from only 11 or 12 flare events, near the 10-flare minimum for the catalog. If the flux errors in the light curve of the most extreme outlier were increased by only $\sim 10$\%, the resulting $p$-value increase turns a one-in-a-million event into an unsurprising one-in-a-thousand occurrence.

Data-processing steps (including outlier rejection) may have contributed to this underestimate of errors. Alternatively non-Gaussianity in the error distribution from processing or intrinsic astrophysics may cause the $p$-value outliers. Either effect is more pronounced when data from a small number of light curves are used to build a composite spectrum; the Central Limit Theorem suggests that composite light curves built from more flare events are less susceptible to these effects.

This assessment of the role of noise estimates in our interpretation of composite light curves highlights the value of obtaining as many post-flare light curves as possible for each source. As more flare events contribute to a composite light curve, the sampling errors on flux uncertainties become smaller. The ``noise'' may also come from micro-flare events --- we chose the stars in our survey precisely because they are active. This possibility is more reason to strive for high $\nflares$. 

\begin{figure}
    \centering
    \centerline{\mbox{\includegraphics[width=0.5\textwidth]{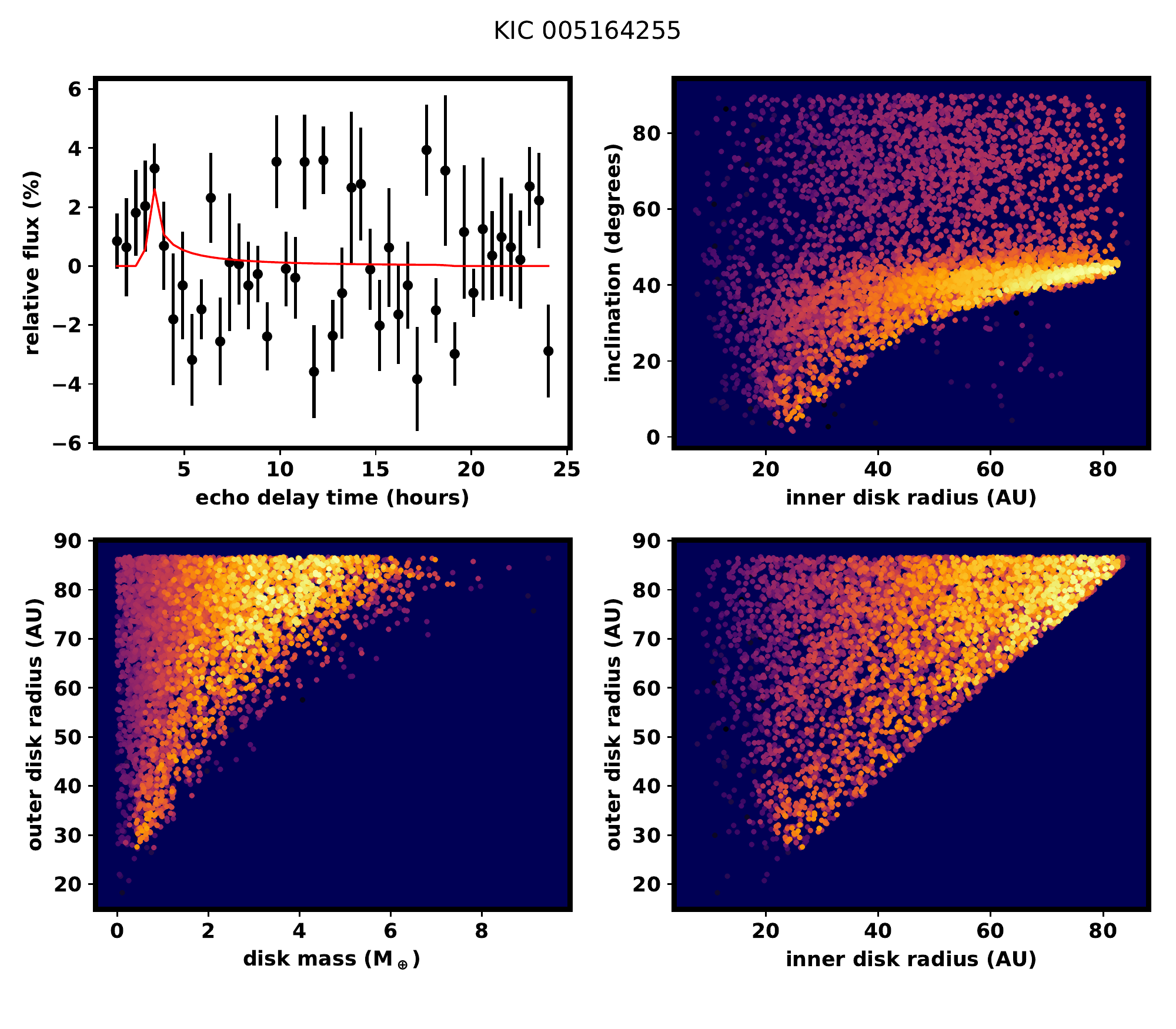}
    \includegraphics[width=0.5\textwidth]{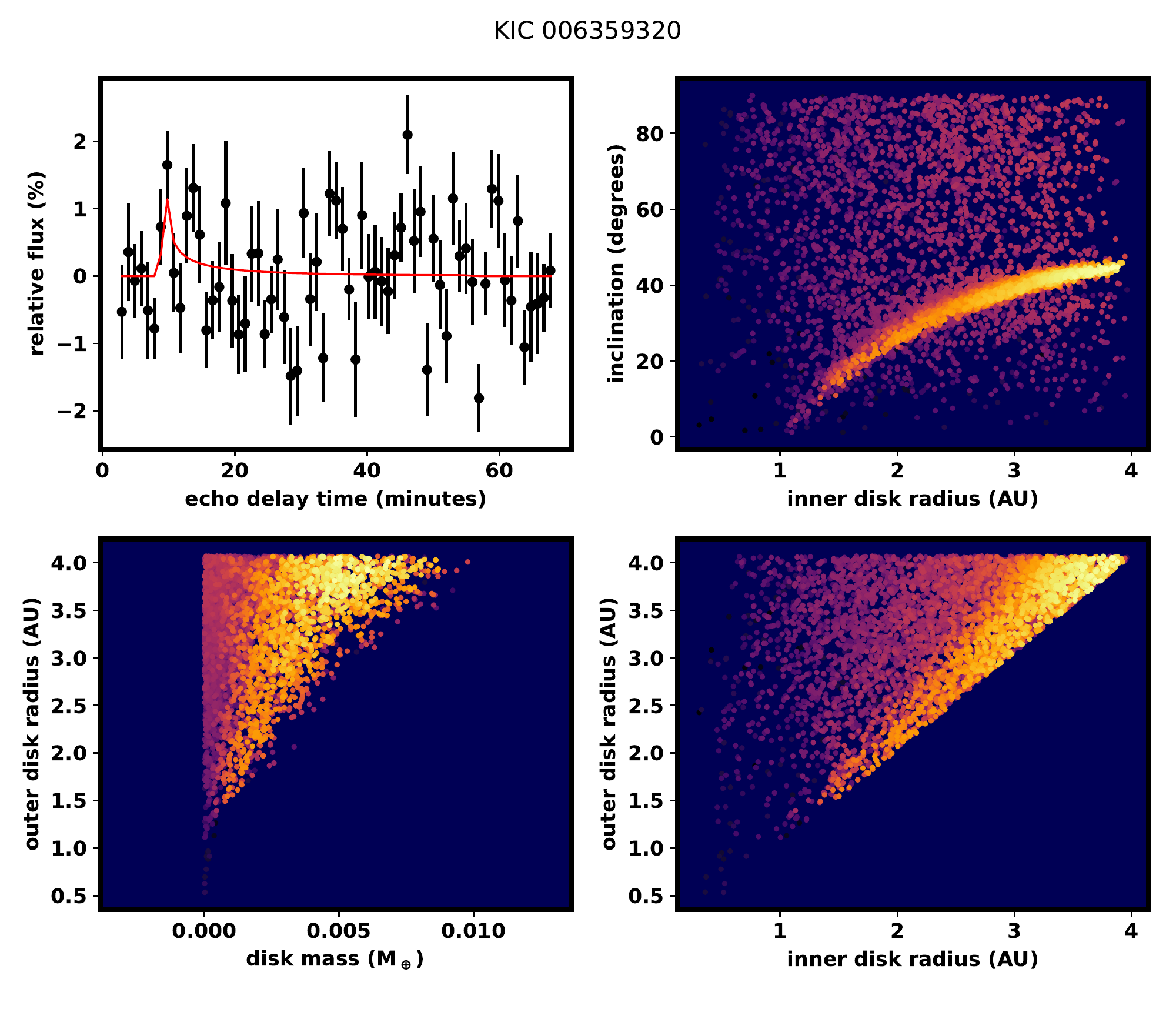}}}
    \caption{\label{fig:outliers} 
    Light curves and MCMC parameter-space samples from two outliers. The light curves depicted here are the most significant outliers from the long-cadence and short cadence data sets. They are both formally inconsistent with zero surplus flux and the level of uncertainties as shown by the 1-$\sigma$ error bars. The best-fit echo curves suggest that we are fitting to the noise: random fluctuations, not an echo signal, appear responsible for the inconsistency.}
\end{figure}

Despite these conclusions, it is worth exploring the outliers for which the no-echo hypothesis is formally rejected. In these cases, large random scatter about the quiescent background, as opposed to a strong coherent echo signal above it, appear to be responsible for the low $p$-values. Figure~\ref{fig:outliers} provides examples. 

No individual composite light curves provide compelling evidence of flare echoes from circumstellar disks around late-type stars.  Despite this result, these light curves offer meaningful disk mass limits, as we describe next.

\subsection{Model fits to composite light curves: likelihood analysis}\label{sec:modelfits}

The composite light curves place formal constraints on disk model parameters, including mass limits (\S\ref{subsec:meth}). We generated MCMC samples of the likelihood function in Equation~(\ref{eq:likelihood}) from fits to each of the 2,361 composite light curves. Figure~\ref{fig:masslim} contains our results (data for this figure are available in the electronic table), illustrating these central messages: 

\begin{figure}
    \centering
    \includegraphics[width=0.7\textwidth]{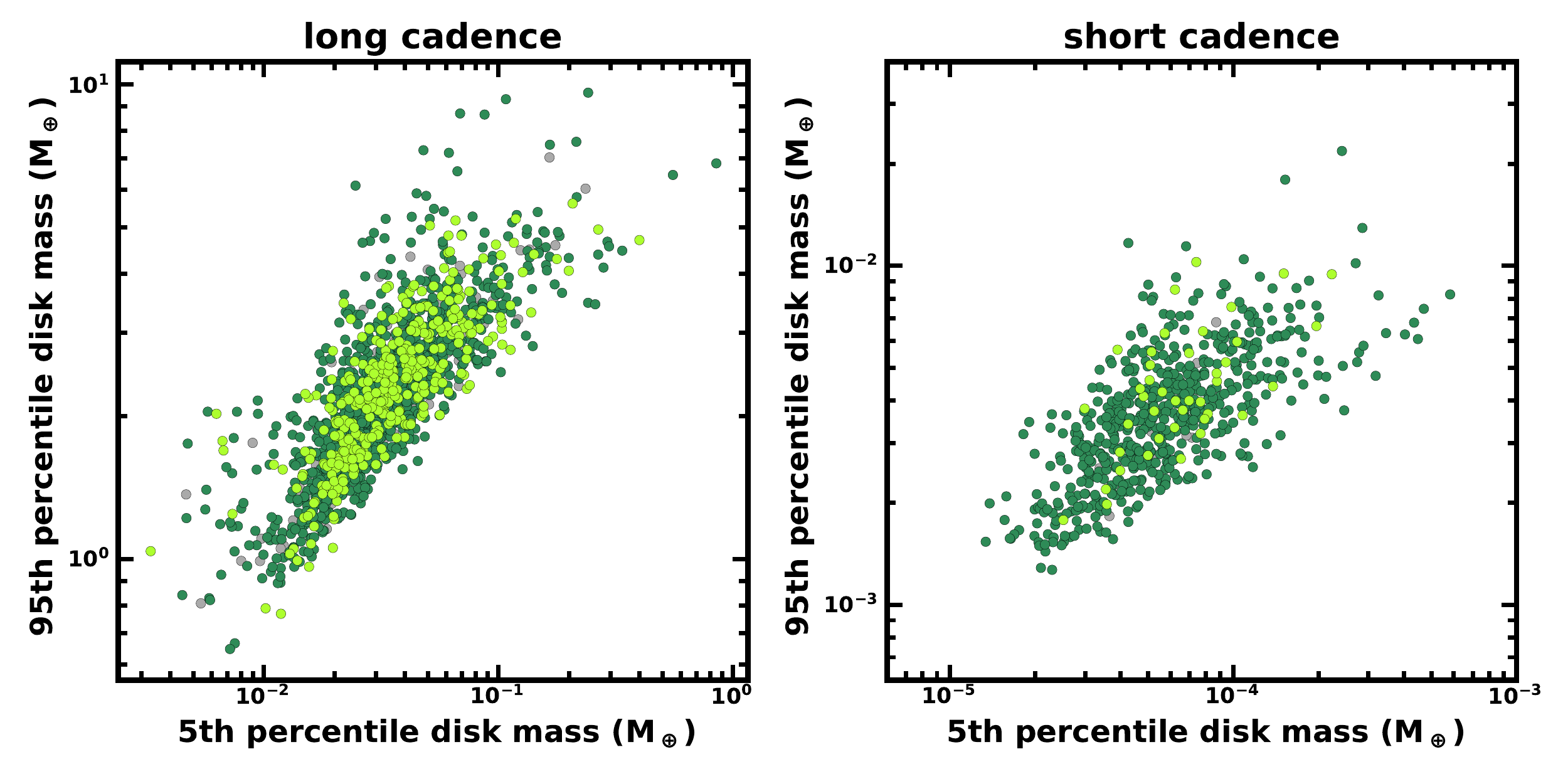}
    \caption{\label{fig:masslim} MCMC mass limits from long-cadence and short-cadence light curves. The shade of the symbols indicate WISE W1-W3 colors; the lighter shade show stars with IR excess (W1-W3$>1$~mag), while the darker shade designates stars with weaker or no IR excess.}
\end{figure}

\begin{itemize}

    \item \textit{Disk masses for exo-Kuiper belts are not well constrained.} With the long cadence data, the formal mass limit for disks at $\sim$7--90~au composed of micron- to millimeter-sized particles is 1--3 Earth masses (Table~\ref{tab:masslim}). A disk with mass in this range is near or over the threshold of being optically thick, unless its scale height is comparable to its radial extent, as in a torus (cf. Eq.~(\ref{eq:mthresh})).

    \item \textit{Mass limits in the terrestrial zone are more stringent.} From the MCMC likelihood sampling with the short-cadence light curves, disks of $\micron$- to mm-sized dust with $\aout \lesssim 4$~au have masses below $\sim$0.01~\Msolar, about a lunar mass. The inferred masses in these cases are consistent with optically thin disk models even when the scale height is modest ($h\sim 0.05$ in Eq.~(\ref{eq:h})).
    
    \item 
    \textit{The IR (W1-W3) color and disk mass limits are weakly correlated.}
    Stars are grouped according to their WISE W1-W3 color to identify objects with IR excess ($0.2\leq$W1-W3$<1$~mag), strong excess (W1-W3$\geq 1.0$~mag), or no IR excess (W1-W3$<0.2$~mag), as in Table~\ref{tab:masslim}. From one-sided Kolmogorov-Smirnoff tests, the stars with strong IR excess have $\mlim$ distributions shifted to higher mass than stars with no excess (95\% confidence for short-cadence data, and 99.9\%\ confidence for long-cadence data). This trend, greater possible disk masses in cases of strong IR excess, is consistent with the idea that when there is dust, there are echoes. However, it may also be consistent with more micro-flares occurring around younger, more active stars that also have more dust.

\end{itemize}

\begin{deluxetable}{lrclll}
\tabletypesize{\footnotesize}
\tablecolumns{6}
\tablewidth{0pt}
\tablecaption{Mass limit summary. \label{tab:masslim}}
\tablehead{\colhead{population} & \colhead{$N_\text{obs}$} & \colhead{$\ain$--$\aout$} & \colhead{median $\mlim$} & \colhead{16th--84th percentiles} & \colhead{ensemble $\mlim$}}
\decimals
\startdata
LC all  &  1734  &  7.4--86.7 au  &  2.194 \Mearth  &  1.558--3.152 \Mearth &  0.128 \Mearth\\
LC no IR excess  &  515  &   \   &  2.169    &  1.516--3.200   &  0.082  \\
LC IR excess  &  747  &   \   &  2.155    &  1.515--3.068 & 0.108 \\
LC strong IR excess  &  385  &   \   &  2.355    &  1.660--3.244  &  0.473 \\
LC G stars  &  846  &   \   &  2.178    &  1.552--3.125    &  0.129 \\
LC K stars  &  384  &   \   &  2.224    &  1.581--3.330  &  0.263 \\
\hline
SC all  &  627  &  0.25--4.07 au  &  3.637 $\times 10^{-3}$ \Mearth  &  2.324--5.664 $\times 10^{-3}$ \Mearth  & 0.182 $\times 10^{-3}$ \Mearth \\
SC no IR excess  &  403  &   \   &  3.647    &  2.227--5.571 & 0.129  \\
SC IR excess  &  170  &   \   &  3.508    &  2.549--5.669 &  0.433 \\
SC strong IR excess  &  41  &   \   &  4.117    &  2.930--6.377  & 0.782 \\
SC G stars  &  293  &   \   &  3.551    &  2.387--5.589   &  0.204 \\
SC K stars  &  147  &   \   &  3.549    &  2.323--5.196   &  0.432 \\
\enddata
\tablecomments{In column one, we label objects according to their level of IR excess --- a possible surrogate for the presence of circumstellar dust. \citet{rizzuto2012}, identify a color range of W1-W3 $\approx 0.2$--0.3~mag that delineates objects with disks from those with no detectable dust. Here, all sources with $0.2~\text{mag} \leq \text{W1-W3} < 1.0$~mag are considered to have an IR excess, while the excess is ``strong'' for the subset of stars with W1-W3 $\geq 1$~mag. Objects with W1-W3 color that is less than 0.2~mag are labeled as having no excess. The second column provides the number of objects that fall into these categories. The next three columns are the radial extents of disk models applied to each set of stars, the median mass limit, and the 16th-84th (68\%) range of $\mlim$ values within each population. The last column gives results for mass limits from the analysis of ensemble-averaged
light curves described in \S\ref{subsec:allstars}.}
\end{deluxetable}

\newpage
\subsection{What are the limits? Artificial echo injection and detection thresholds}\label{subsec:artifact}

To evaluate the feasibility of detecting disk echoes in this sample of light curves, we perform artificial echo injection and detection on the 1,734 long cadence and 627 short cadence candidates. Echoes are injected assuming an observable disk under perfectly ideal conditions: an optically thin disk with isotropic scattering in a face-on orientation. This hypothetical disk is also assumed to be a Kuiper Belt analog, with an inner radius $a_{in} = 30$ AU and an outer radius $a_{out} = 50$~au, a minimum dust grain radius of 1 micron ($r_{min} = 1 \mu m$), a monodisperse dust particle distribution, and a dust particle density $\rho = 2 g/cm^3$. Finally, we assume that the artificial echoes represent the largest amount of possible reflected light from a disk --- while each flare will illuminate only half of the disk, the total amount of light gathered amounts to illumination of the total surface area of the disk. These assumptions allow us to relate detection limits to the disk’s mass, as in Equation~(\ref{eq:ftot}). The artificial echoes are injected into light curves at a post-flare index consistent with an inner radius of 30 AU – about the orbit of Neptune. The analysis is performed with the additional assumption that all flares in this sample of stars are unresolved delta flares that last a single time bin, and that the observable portion of the echoes also last a single time bin at a small percentage of the peak flare flux value. 

\begin{figure}
    \centering
    \includegraphics[scale=0.7]{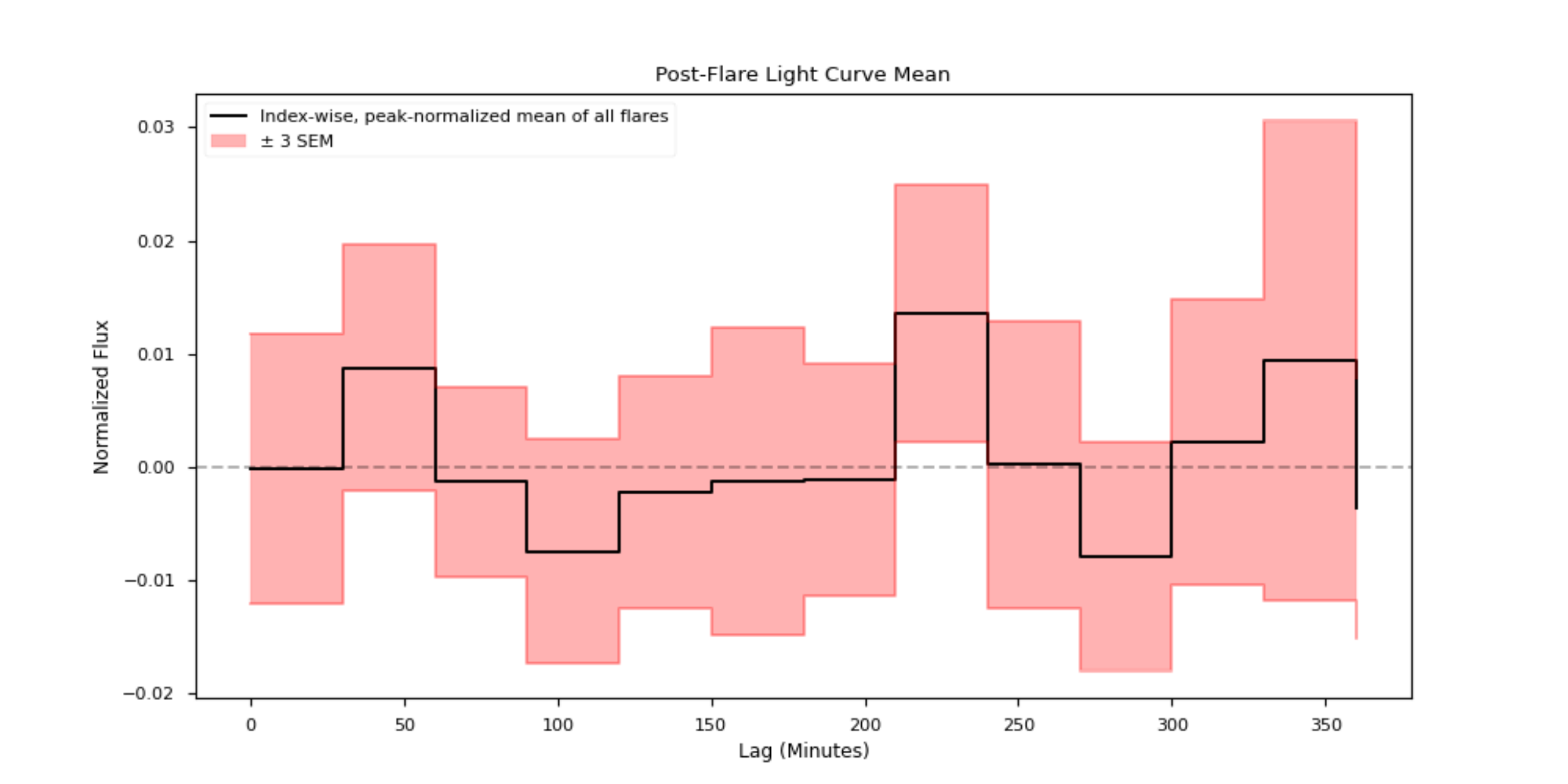}
    \caption{Post-flare normalized light curve for KIC 2834053. The black curve shows the mean of all flares after normalization by the peak flare flux value, and the red shaded area represents $3\frac{\sigma}{\sqrt{n}}$ on either side of the mean. This particular case showcases the detection of an artificially injected echo at 1\%\ echo strength, seen at a lag value of approximately 210 minutes.}
    \label{fig:postflarelc}
\end{figure}

The data are prepared in a manner consistent with the steps outlined in \S\ref{subsec:meth} --– flares are treated as unresolved and normalized by the peak flare flux value, and the average flare light curve from all flares available in the star at a 6-sigma threshold is used. During flare processing, 10 long cadence and 23 short cadence stars are removed from the sample because they lacked enough 6-sigma flares to meaningfully evaluate artificial echoes. To produce detection criteria, we evaluate the standard error of the mean at each index. Echoes are considered ``detected'' if the following inequality held true at the lag index of interest:
\begin{equation}\label{eq:fake_echo}
\mu - 3\frac{\sigma}{\sqrt{N}} > 0 ,
\end{equation}
where $\mu$ is the composite light curve mean at the lag index of interest, $\sigma$ is the standard deviation at that index, and $N$ is the total number of flares used. The detection threshold is chosen to be three times the standard error of the mean --- a 99.7\% confidence interval. This detection process is illustrated in Figure~(\ref{fig:postflarelc}), where it is shown being applied to the star KIC 2834053, a rotationally variable K star with a high level of flare activity. This analysis allows us to identify detection limits, as well as test false-positive rates for the bins where no echo is injected.  

\begin{figure}
    \centering
    \includegraphics[scale=0.85]{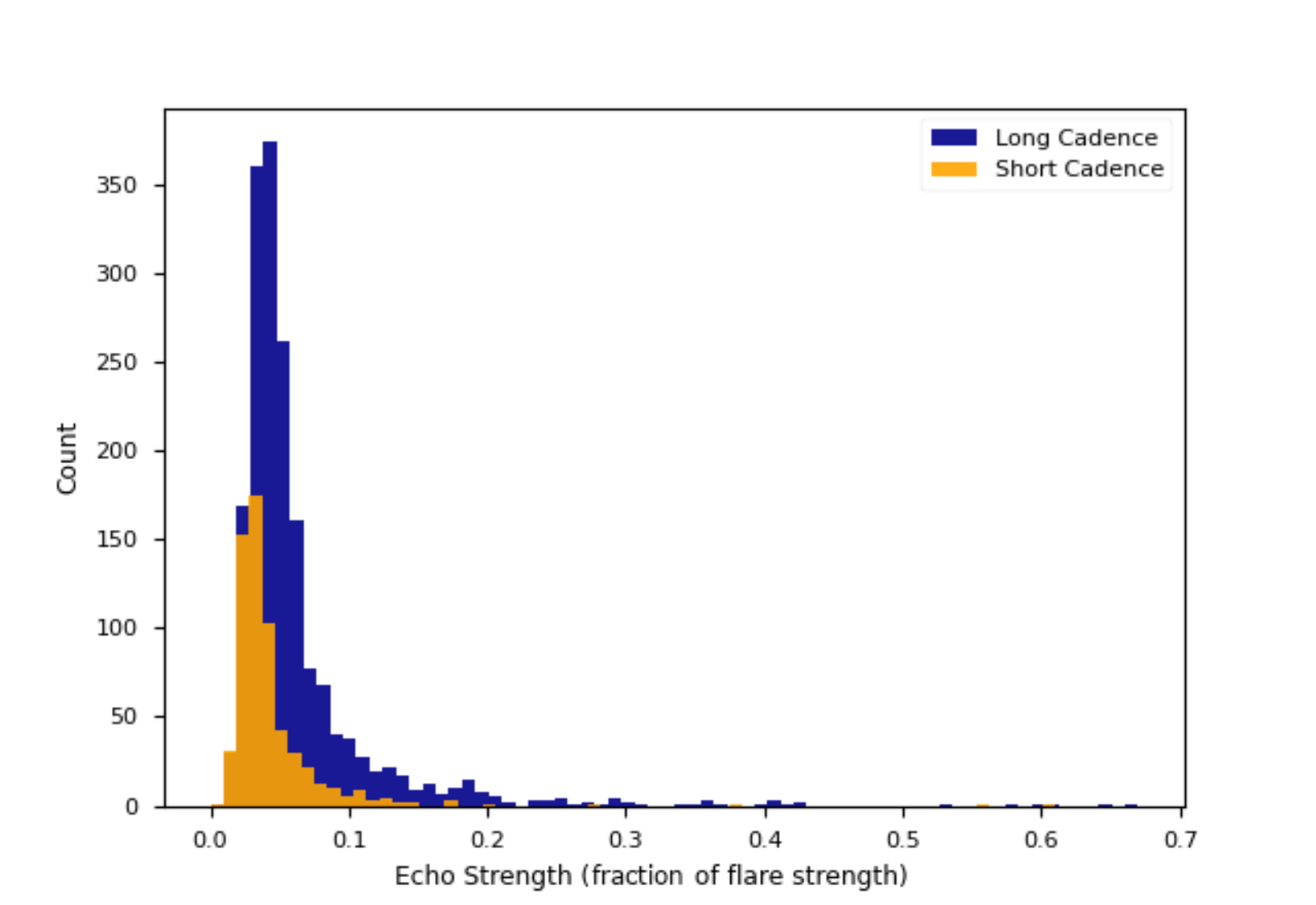}
    \caption{Distribution of lowest detectable echo strengths {at 99.7\%\ confidence} for the full processed sample of 1,774 long cadence stars and 611 short cadence stars. 
    }
    \label{fig:lde_hist}
\end{figure}

Results closely follow the theoretical predictions outlined in \S\ref{sec:bg}. In the optically thin limit, disks with higher masses reflect higher amounts of light. Additionally, this analysis reveals that a vast majority (67\% of long cadence stars, and 80\% of short cadence stars) of our top candidates may have detectable echoes -- or at least statistically interesting data points -- at or below a 5\%\ echo strength threshold. Detectable echo strengths range from 1\%\ to 67\%. As discussed in \S\ref{subsec:disk}, the Kuiper Belt can realistically reflect 5\%\ of the total light from the Sun, implying that several candidates in this sample could have observable Kuiper Belt-like disks under ideal conditions. Further, 30 long cadence and 31 short cadence stars were found to have detectable synthetic echoes at 1\% echo strength, which is sensitive enough to prompt further investigation. The full range of echo strengths found in this assessment is shown in Figure~(\ref{fig:lde_hist}). Using Equation~(\ref{eq:ftot}), we can also estimate the mass of these theoretical disks. For example, KIC 2834053 has an echo detection ``floor'' of 1\%\ echo strength, which corresponds to a disk with mass $M_{disk} = 0.001875  \Mearth$ -- about 20\% of the mass of the Kuiper Belt in small dust grains. Estimated disk masses in this sample range from $0.001875 \Mearth$ to $0.126 \Mearth$ in micron-sized dust.

While false positives are one of the chief concerns of this work, the possibility of false negatives remains a real concern, even at 99.7\%\ confidence. To illustrate, we let the probability of detection be $P(D) = 0.997$ for a given echo in a single star. The probability of missing that detection --- or a false negative --- is then $P(M) = 1 - P(D) = 0.003$. With 1784 stars in the database, the probability of finding an echo in each star becomes $P(D) = 0.997^{1784} = 0.0047$ and the probability of at least one false negative is $P(M) = 1 - P(D) = 0.9953$. There is an almost certain chance that at least one detection would be missed in this sample: With roughly 1000 sources, there is a 95\%\ chance of at least one false negative.

While this proof-of-concept technique is simple, and requires many assumptions, it is useful to apply limits on the detectability of unresolved echoes in each light curve. Furthermore, placing simplest-case limits on the disk mass for each of these candidates helps give physical expectations for further study. Real disks almost certainly will not present such ideal observing conditions, but techniques such as this one lay the groundwork for more advanced echo detection methods using the data in this sample, and provide a frame of reference for those results.

\newpage
\subsection{Ensemble light curves}\label{subsec:allstars}

\begin{figure}
    \centerline{\includegraphics[width=0.95\textwidth]{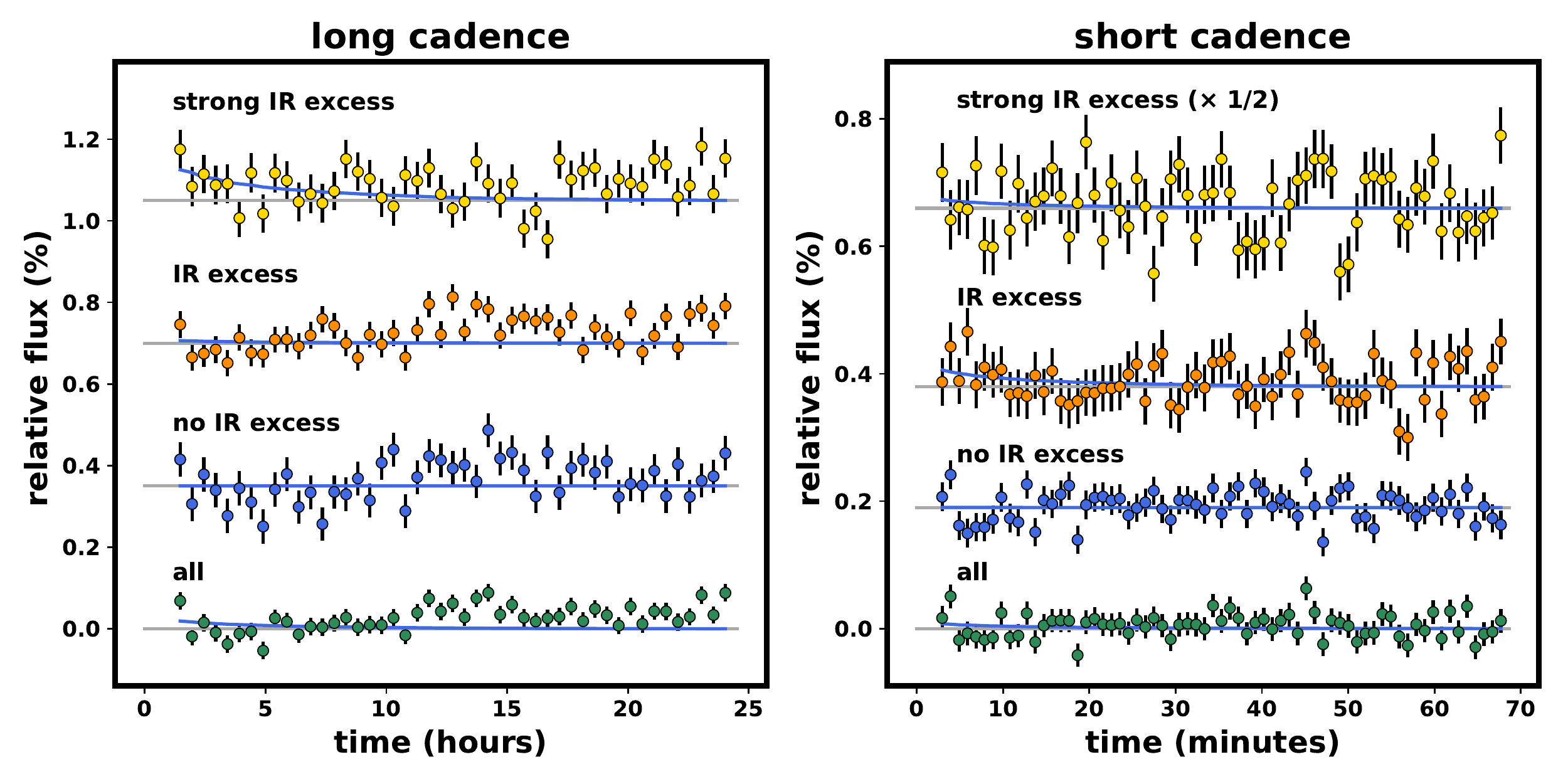}}
    \caption{\label{fig:allstars}The multi-star ``ensemble'' light curves for long-cadence and short cadence data, grouped by level of IR excess. The bottom set of (green) points in each panel are the combined light curves from 1,734 stars and 627 stars respectively. The second set of points (in blue) are sources with no IR excess (W1-W3$<$0.2~mag; see Table~\ref{tab:masslim}), and the set above that (in orange) have modest IR excess (between 0.2 and 1.0~mag), while the topmost set are stars with strong IR excess ($>$1.0~mag). The light grey lines show the no-echo expectation, while the blue curves give the best-fit models for the average of randomly oriented disks.} 
\end{figure}

The composite light curve of any one star in our sample has a noise level that exceeds reasonable echo signals from circumstellar dust. However, if dusty disks are common around active stars, the combined faint echoes from many stars would produce persistent surplus signal in post-flare light curves. To explore this possibility, we first combine all 1,734 long-cadence light curves into a single ensemble-averaged one, and do the same for all 627 short-cadence composite light curves. Figure~\ref{fig:allstars} shows the results: two ensemble (``all-star'') light curves, built from a variance-weighted average of mean fluxes in time bins. The errors in these flux bins are a few parts in 10$^{-4}$. Importantly, the average quiescent starlight is constrained to within 0.1\%\ of the flare peaks. This reduction in noise, achieved with Kepler data, bodes well for observing light echoes in extensive observations of single stars from observatories equipped to perform such a task.

To assess the possibilities for constraining disk masses with the multi-star data in the panels of Figure~\ref{fig:allstars}, we build model light curves for echoes from ensembles of disks at random orientations. In our numerical models, we implement averages by sums of disk models on a grid of orientations weighted according to a $\sin\inc$ distribution. The model echo profiles generally have strong prompt signal, since geometry and forward-scattering by dust generally favor echo contributions from the forward part of inclined disks. The averaged echo starts at a high level immediately after flares and decays exponentially thereafter. The \mysoft{emcee} code samples the likelihood distribution of fits to data with these models, assuming that there are inner and outer disk radii that characterize the dust distribution averaged over all sources. The output is a single best-fit inner radius, outer radius and average disk mass. 

The results are tight constraints on the average mass of disks composed of micron- to millimeter-sized dust: The limit for the long-cadence catalog is 0.13~\Mearth\ (95\%\ of MCMC samples) at distances of 10--90~au, and below $3\times 10^{-4}$~\Mearth\ within 4~au for the short-cadence light curves (see Table~\ref{tab:masslim}). These limits hold for all average dust distributions that have surface density profiles with the form of Equation~(\ref{eq:sigma}). The dust around any one star may be ring-like or spread out; the main assumption is only that the dust surface density decreases with radius when averaged over many stars.

Grouping stars according to whether they show an IR excess yields an interesting, and perhaps significant, trend; the greater the IR excess, the greater the mass limit derived from the MCMC samples. The effect is more pronounced in the long-cadence data. We expect that IR excess is an indicator of circumstellar dust, and the limits indicated in our analysis are comparable to masses in observed debris disks \citep[e.g.,][]{su2015}. These results are still noise-dominated, but trend in a tantalizing way consistent with the presences of disk echoes.

Alternatively, we group stars by spectral type, with the rationale that the dust distribution around stars of similar type may be similar as well, both in terms of radial distribution and total mass. For example, in the short-cadence data, disks are resolved within 4~au, yet the snow line around K stars lies within an AU, and for F stars is well outside 4~au. If the mass reservoir for dusty debris is discontinuous inside the snow line, then the assumption of a decreasing ensemble-averaged surface density would have to be reassessed, depending on the mix of spectral types. Fortunately, the overwhelming majority of stars in our catalogs (G and K types) have snow lines that are well interior to the disks resolved by the long-cadence data.

\begin{figure}
    \centerline{\includegraphics[width=0.95\textwidth]{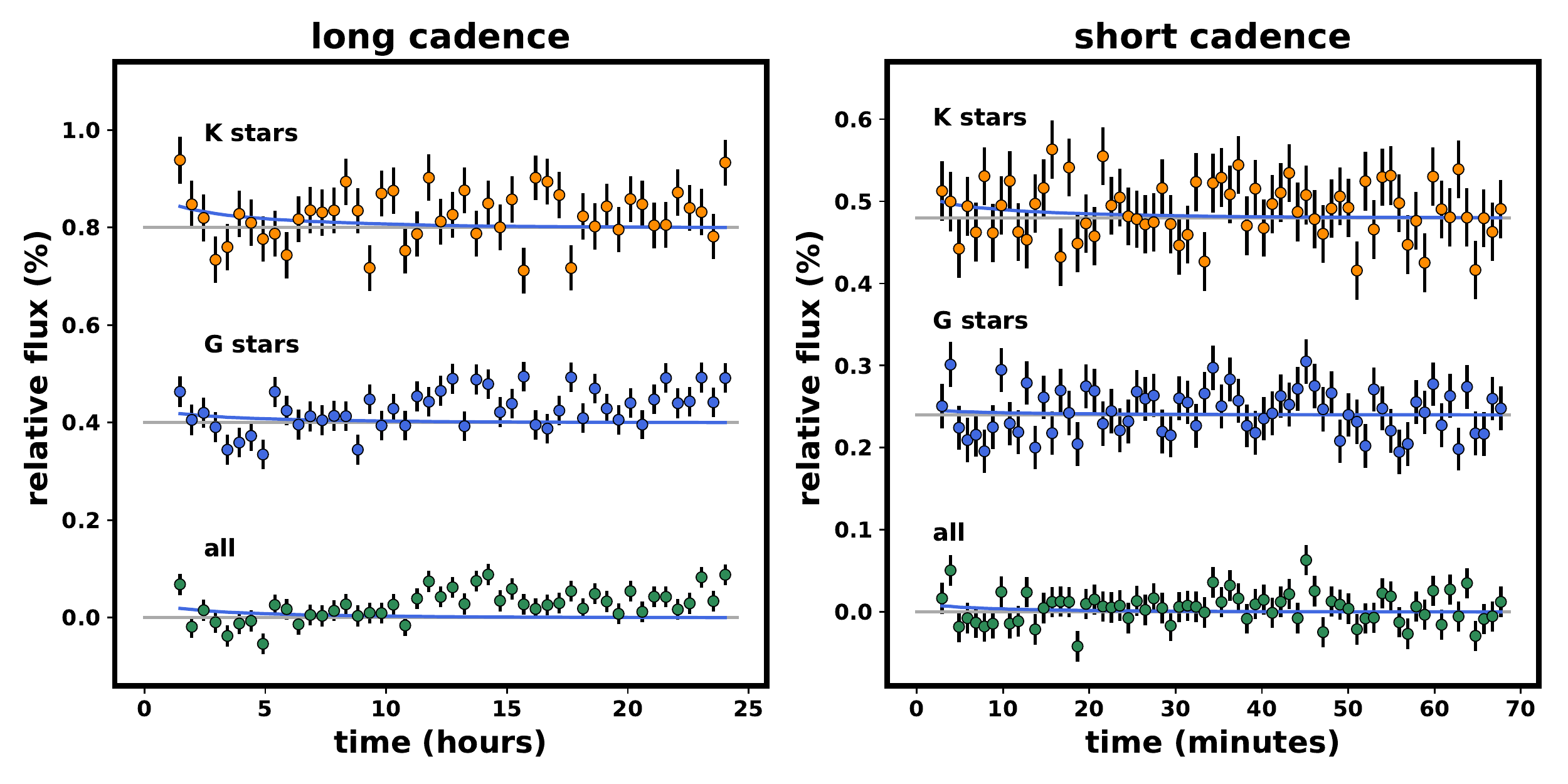}}
    \caption{\label{fig:allGK}The G-star and K-star ensemble light curves, similar to the previous figure. The ``all-stars'' data are displayed in the lower part of each panel, while G stars (846 and 293 objects in the long- and short-cadence catalogs, respectively) are shown in the middle. The upper part of each panel gives results from  K stars (384 and 147 objects). The grey curve is the expected light curve with no echo, and the blue curves are model light curves representing the ensemble average of stars with randomly oriented disks.}
\end{figure}

Motivated by these concerns, we derive ensemble light curves for G and K stars separately, showing results in Figure~\ref{fig:allGK}; Table~\ref{tab:masslim} provides numerical mass limits. From the MCMC model samples, the disk mass limits on K stars are less stringent than on G stars; 0.26~\Mearth\ versus 0.13~\Mearth\ in the long-cadence data, and 4\%\ of a Lunar mass versus about 2\% in the short-cadence data. Noise in the ensemble light curves, which increases with decreasing number of light curves (and stars), is likely responsible for these results; there are roughly half as many K stars as G stars in the each of the long-cadence and short-cadence catalogs. We note with caution that in the long-cadence catalog, there are as many K stars as there are stars of any type with strong IR excess, yet the latter admit a much higher ensemble-averaged disk mass (0.26~\Mearth\ compared with 0.5~\Mearth).

While our analysis of ensemble light curves has shortcomings, including the assumed form of the averaged disk surface density,  the ability to produce limits that are comparable to observed disk masses speaks to the potential of the method described here.

\newpage
\section{Discussion}\label{sec:discuss}

\subsection{Signal processing issues}

The work presented here is a challenging problem in signal-processing, with the goal of detecting faint echoes of stellar flares from an active star with a circumstellar disk. As illustrated in \S\ref{sec:bg} with simulated data  (Fig.~\ref{fig:echofake}), uncertainties in flux must be a fraction of a percent relative to the impulse flare amplitude to extract astrophysically relevant estimates of disk parameters from echoes. The simulated example also highlights a key element of our strategy, to generate co-added, composite light curves. This step solves two problems: First, individual flares illuminate different parts of the disk and have uniquely shaped echo profiles. Co-adding $O(10)$ flares is sufficient to build out an average echo profile that can be easily modeled. The second issue is that with real data, we expect to be limited by noise. Co-added flares helps mitigate that problem.

We have these signal processing issues in mind when approaching data from the Kepler mission. We select impulsive flares that stand out relative to the background noise, not just the total quiescent flux. We also require ten or more unresolved, impulsive flare events. The resulting set of light curves is thus tuned for the problem of echo detection. 

Even though our selection of flare events is performed at high signal to noise, our flare detection algorithm may be susceptible to false positives. That we require ten such events over the full course of observations of any one star may mitigate the problem: The events are ``repeatable'', arguing for an astrophysical origin. Still, \citet{jackma2021} analyzed short-cadence data and determined that roughly 7\%\ of flare event detections are false positives from confusion with a faint, nearby star. The impact on the results presented here is that sometimes we may be looking for echoes around the wrong source. This problem dilutes signal, but does not trigger false detections; indeed, if all flares are from the neighboring source, any detected echoes would be, as well.

There remain other signal processing issues. One is possible noise from microflares. Our effort to mitigate this is to implement a stiff, multi-step outlier rejection protocol as described in \S\ref{sec:kep}. The idea is that echoes are typically long-lived and below the noise in individual light curves, so we can remove unwanted events by rejecting single-bin outliers.  

The second key processing element is the removal of long-term signals associated with the star, for example from the presence of starspots that modulate the stellar background as the star's rotation brings them in and out of an observer's field of view. This time scale is typically days to weeks. Still, echoes from an extended exo-Kuiper belt can be as long lasting as a day, so filtering out the stellar modulations to isolate an echo can require a balance. We have explored different possibilities for this high-pass filtering and made our choice; see Appendix~\ref{appx:sensitivity} for details on the interplay between the filtering window and our overall sensitivity. 

Finally, we have made a key simplifying assumption that impulsive flares have the shape of the time bin that they are observed in, and that they are coincident with the time index of that bin. In reality, the flares are unresolved and may have taken place at any point within the time bin. There is some uncertainty as to how the echo of the flare maps into subsequent bins. This effect is not relevant for our work here, but may impact future efforts if the flux sensitivity is high. Appendix~\ref{appx:unresolved} discusses how unresolved flares fold into the echo detection process. 

After the application of these signal processing steps, we can assess the success of the overall method by looking at ensemble light curves. These catalog averages illustrate what is possible with data from $10^4$ flare events at the sensitivity of Kepler. At long cadence, the all-stars light curve has contributions from over 64,000 flares, while at short cadence, the total number of events is more than 34,000; both catalogs yield errors in flux, relative to flare maxima, to within a fraction of a percent, and constrain the mean quiescent stellar background at a similar level. These outcomes bode well for future high-sensitivity studies.

One unexpected feature in the long-cadence light curve is the steady, significant trend toward higher flux with increasing lag. 
That the same phenomenon is not apparent in the short-cadence data may argue for an astrophysical origin tied to a slow process like stellar rotation. Possibly the trend reveals a tendency for the stellar quiescent light to be suppressed slightly during observed flare events owing to their association with starspots --- presumably the loci of flares on the stellar surface in visible light. During flares, active regions are assured to be present, while at other times, stellar rotation may remove star spots from our view of the stellar disk \citep[e.g.,][]{davenport2015}. If this speculation is correct, we expect a similar analysis in UV light to have a decrease in flux with time following flare events, since star spots are bright in the ultraviolet. 

\subsection{Limitations of the echo modeling}

In addition to signal processing issues, there are other considerations and caveats to this work related to the modeling. We have made simple approximations for dust and disk properties (\S\ref{sec:bg}), which may relaxed and incorporated as fitting parameters. Also, we have idealized the nature of flares as impulsive and localized on the stellar surface, so that each flare illuminates only half the circumstellar disk. Flares may be more extended in time and space, protruding to large distances from the stellar surface; then they illuminate more of the disk, changing the shape of the expected composite echo profiles (Eqs.~(\ref{eq:echogrid}) and (\ref{eq:fmulti}). Other limitations of our modeling include the assumptions that the stellar radius is small so that flares fire off at the origin. More realistic configurations are necessary for dust that is close to the host star. Finally the dusty disk structure may be much more complex, and may not even be axisymmetric. Astrophysical disks can be carved up by embedded planets or shepherded on eccentric orbits by gas giants. However, for this feasibility study with data dominated by noise, these limitations do not affect our overall conclusions.

\subsection{Sample selection and flare statistics}

The catalogs presented here, subsets of the Kepler Input Catalog, offer limited information on the demographics of flare stars. Our focus on normalized, unresolved flares means that this work is not designed to contribute to studies of detailed flare properties \citep[e.g.][]{hawley2014}. For broader demographic assessments such as the dependence of flare event rates on spectral type \citep{balona2015, yang2019}, our selection of stars is also problematic. We consider flare amplitudes only as they relate to instrumental noise and counting statistics, and include stars only after outlier rejection culls extreme flare events. Despite these differences, the spectral type of the active stars in our sample (Fig.~\ref{fig:spectral_type}) offer no surprises when compared with other studies.

Where our catalogs might offer something unique from a demographic standpoint is the connection with WISE IR excess. In the long-cadence catalog, two thirds of the sources show IR excess (W1-W3$>$0.2~mag), and more than half of these stars have strong excess (W1-W3$>$1~mag). That there should be a correlation between activity and circumstellar dust is reasonable, since both are associated with younger stellar ages. Since the short-cadence catalog corresponds to unresolved flares that have significantly less energy than in the long-cadence catalog, it contains fewer active stars in terms of flare energetics. The short-cadence catalog has fewer sources with IR excess, too, with only a third of its members having w1-W3 color above 0.2~mag. The short-cadence stars are thus likely to be an older population than the long-cadence stars.

The implication for echo searches is that younger stars make for more promising sources of echoes because of their greater activity and prevalence of circumstellar dust. 

\subsection{Interpretation of the disk mass limits}

Finally, we discuss the interpretation of mass limits in the composite and all-star light curves. The composite light curves in the long-cadence data have noise levels that admit disk models with mass large enough that they are no longer optically thin in some cases. Thus, the limits are often of limited utility. The short-cadence catalog is better in this regard, although the masses are still close to the threshold where the optical thickness may be substantial (Eq.~(\ref{eq:mthresh})). Even so, the mass limits --- roughly an Earth mass and fraction of a Lunar mass for long-cadence and short-cadence data, respectively --- are too large to be astrophysically useful: Observations of micron- to millimeter-sized dust in circumstellar disks indicate that we may expect only a fraction of an Earth mass in the outer regions of planetary systems \citep[e.g.,][]{su2015, marino2018} where our long cadence analysis applies. In the terrestrial zone, where the short-cadence analysis is most sensitive, observation of warm dust suggest the presence of only a few percent of a Lunar mass \citep[e.g.][]{weinberger2011, fujiwara2012}.

Stars that are grouped according to their WISE W1-W3 color, a surrogate for the presence of circumstellar dust, show that disk mass limits $\mlim$ increase with IR excess. This trend persists in the long-cadence data, the short-cadence data, the composite light curves, and the ensemble light curves. This correlation might reflect counting statistics, since higher uncertainty in fluxes admits higher limits on echo strength. However it persists in the ensemble light curves that have the least amount of noise, hinting that this trend might have an astrophysical basis.

Unfortunately, models for inclination-averaged light curves peak at zero lag when there is forward scattering from the dust. At short lag, there may be confusion with the exponential decay of individual flares, so we cannot distinguish between echo and flare {with a photometric measurement}. Nonetheless, model fitting can provide stringent mass limits from ensemble light curves; as in Table~\ref{tab:masslim}, these limits are comparable to the dust mass in observed debris disks.

\subsection{The future for disk echoes}

Kepler was not designed to detect echoes from disks, so our weak limits on disk masses are unsurprising.  First, Kepler detects in a broad photometric band where echoes must compete with significant background; a U-band observatory would have significantly higher signal-to-quiescent background.\citep[]{bromley1992} Second, flare frequency scales with flare intensity \citep[e.g.][]{davenport2016b, vida2018}: more flares are detected per quarter in the short cadence data than in the long cadence data, so higher cadence observatories will acquire far more flares.  The Transiting Exoplanet Survey Satellite \citep[TESS]{ricker2015} is observing select bright stars at a two-minute cadence and may provide better opportunities for echo detection than Kepler's long-cadence campaign. Third, the Kepler satellite's collecting area of 0.7~m$^2$ is tuned for identifying transiting planets, but faint echoes require larger telescopes with greater sensitivity; the Nancy Grace Roman Space Telescope and LUVIOR are examples. Finally, big data measurements have big data problems: given a 99.7\%\ threshold and 40+ bins per star, the likelihood of a false positive somewhere in the composite light curve is already on the order of 10\%\, making a confirmed detection a challenging task even with a statistically significant outlier; applying this to thousand-plus star catalogs and the number of false positives is expected to be considerable, which is consistent with our results.  We performed additional statistical analyses, including bootstrap resampling and jackknife measurements, and while several of our measured `hits' survived all tests, we ultimately decided not to declare any single event an echo detection.  For a mission designed to detect disk echoes, a predicted false-positive rate of $<1\%$ per star (approximately 4-sigma in this experiment) at meaningful echo magnitudes, which are also on the order of $<1\%$, will be required.

Even with a large, fast, sensitive, persistent U-band observatory, one additional concern must be addressed: microflares.  There will always be stellar activity just below the noise floor, no matter how low the floor goes.  We handled this problem through leave-1-out and leave-2-out outlier rejection, but more generalized ``leave-p-out'' measurements will become necessary when flare catalogs are large enough.  However, an additional discriminator could have considerable value, like from a multispectral measurement.  Unfortunately, very limited multispectral data is available from flares, and the data that exists is from a handful of events on a handful of stars,\citep[e.g.][]{kowalski2019} far from a survey.  As we look forward to data from new ultraviolet spectroscopic observatories in coming years \citep[e.g.][]{flemming2018}, we postulate that flares will produce sufficiently unique thermal-temporal curves that multidimensional templates can be used for echo detection with sufficient confidence to confirm a disk detection.  For survey applications, a UV imaging spectrograph in the style of an objective-prism-spectrograph, though perhaps equipped with a much lighter grating for dispersion, would likely serve this cause.

\section{Conclusion}\label{sec:conclude}

Dusty debris disks are an integral part of all evolving planetary systems, the signposts of on-going planet formation \citep{kb2002}. While containing only a fraction of the mass of observed exoplanets, debris disks are orders of magnitude brighter in reflected light. Here, we follow \citet{gaidos1994} in considering the possibility that dusty disks produce light echoes of stellar flares. Through echo detection, we can identify the presence of circumstellar dust, even when direct imaging of dusty disks is not possible. This effort, to resolve disks in the time domain, complements spectroscopic methods that identify dusty disks from the excess IR emission associated with their thermal radiation. 

Our echo-detection strategy is to adopt simple models for dust properties (size distribution, density and phase function) and the geometry of circumstellar disks (radial extent and surface density) that are guided by observations \citep[e.g.,][]{hughes2018} and theoretical considerations \citep[e.g.,][]{kb2004, kb2008, kb2010}. 
Numerical calculations based on ray tracing then generate echoes as they would appear in post-flare light curves, both individually, when flares illuminate only part of the disk, and as co-added composites. Fits to data with these theoretical light curves provide constraints on disk parameters including radial extent, inclination, and the reflective surface area of the dust. With reasonable choices for dust properties, the fitting procedure offers limits on the mass in micron- to millimeter-sized grains that are responsible for nearly all of the reflected starlight. 

We apply this method to stars in the Kepler Input Catalog, mining the multi-year light curves from NASA's Kepler mission for evidence of multiple impulsive flares, events that are too short-lived to be resolved at the cadence of the observations. The result is a list of over 2,200 stars and roughly 123,000 flare events. Down-selection to ensure at least 10 isolated flare events per star yields a final list of 2,137 stars and 98,834 flares. Working separately with long-cadence and short-cadence data, we build composite light curves formed by co-adding individual light-curve segments, and then perform model fits and likelihood estimations with MCMC sampling to constrain disk properties, including dust mass. 

Our main result is a demonstration that echo analysis of post-flare light curves can place astrophysically relevant mass limits on debris disks. In this preliminary work with the Kepler long-cadence data, we constrain individual dusty disks to a few Earth masses of micron- to millimeter-size dust at orbital distances similar to the Sun's Kuiper belt. The disk masses in this case approach or exceed the limits of our modeling, as these high-mass disks may be optically thick. For comparison, observations suggest that some stars host as much as a fraction of an Earth mass in micron- to millimeter-sized dust in the outer regions of their planetary system \citep[e.g.,][]{su2015, marino2018}.

Constraints from short-cadence data limit the amount of micron- to millimeter-sized dust in the terrestrial zone to roughly $4\times 10^{-3}$~\Mearth. about 0.3~Lunar masses. This constraint is comparatively stringent because dust on close-in orbits scatters a greater fraction of the starlight. Thus, much less close-in dust is needed to account for the same flux level as more distant debris. For reference, the typical mass in millimeter- to micron-sized dust in warm debris disks is roughly $10^{-4}$~\Mearth \citep[e.g.][]{weinberger2011, fujiwara2012}.

Our analysis of the Kepler data is unable to place useful constraints on optically thin debris disks around individual stars because the noise level in the Kepler light curves is too high. However, increased sensitivity in the composite light curves would allow for meaningful mass limits. To achieve greater sensitivity with the Kepler data, we generate catalog-averaged light curves, for which the uncertainty in the mean flux relative to the peak amplitude of flares is low ($\sim 10^{-4}$, as compared with $\sim 0.01$ for individual stars). Using model fits with inclination-averaged composite light curves, we limit the catalog-averaged mass in micron- to millimeter-sized dust to within $\sim 10$\%\ of an Earth mass in exo-Kuiper belts, and $\sim 10$\% of a Lunar mass in the terrestrial zone.  These results demonstrate that increasing the flux sensitivity by a factor of 10--100 in a Kepler-like mission would allow the echo-detection method to bear fruit for individual stars. Otherwise observations of one star with the sensitivity of Kepler would need to run for centuries in order to place stringent limits on disk mass. 

We also consider whether the active stars in this study are likely to have circumstellar dust, as determined from WISE W1-W3 color excess. We associate WISE sources with KIC stars using sky positions; while source confusion and  contamination from extragalactic objects in WISE are a possibility, cross-checks with Gaia DR2 stars suggest that these problems are not prevalent. A search for echoes in light curves from stars with a strong IR excess (W1-W3$>1$) yields disk mass limits that tend to be larger than stars with no IR excess (W1-W3$<0.1$). An intriguing explanation for this trend is that echoes from debris disks are contributing to the light curves of the sources with IR excess. Noise in the Kepler light curves, perhaps from microflaring, seems more likely, if stellar activity and IR excess were connected, for example, by stellar age. Future observing campaigns with greater sensitivity will provide new insight, as they realize the potential of time-domain astronomy for capturing details of debris disks and other sources of echoes.

\acknowledgements

We thank an anonymous referee for suggestions and ideas that improved this work and its presentation. 
We are also grateful to Jeff Arata and Nanohmics, Inc, along with the Center for High-Performance Computing at the University of Utah, for information-technology support for this project. We acknowledge NASA for generous support through the NIAC program, grant 80NSSC18K0041. 
This research has made use of the Simbad database \citep{simbad2000} and VizieR catalogue access tool \citep[for the original description, see \citealt{vizierdoi}]{vizier2000} operated at CDS, Strasbourg, France.
This work also has made use of data from the European Space Agency (ESA) mission {\it Gaia} (\url{https://www.cosmos.esa.int/gaia}), processed by the {\it Gaia} Data Processing and Analysis Consortium (DPAC, \url{https://www.cosmos.esa.int/web/gaia/dpac/consortium}). Funding for the DPAC has been provided by national institutions, in particular the institutions participating in the {\it Gaia} Multilateral Agreement. This publication also makes use of data products from the Wide-field Infrared Survey Explorer, which is a joint project of the University of California, Los Angeles, and the Jet Propulsion Laboratory/California Institute of Technology, funded by the National Aeronautics and Space Administration.

\software{Lightkurve \citep[v1.11.2][]{lightkurve2018}), Scipy \citep{scipy2001}, Emcee \citep{foreman-mackey2013}}

\appendix 
\section{echoes from optically thick disks}\label{appx:tau}

The main focus of \S\ref{sec:bg} is on optically thin disks. Here, we generalize to include optically thick disks. The optical depth along radial lines of sight from the star through the disk is
\begin{equation}\label{eq:tauopt}
    \tau_\text{opt} =  \int_\ain^\aout \frac{\Sigma}{2 H \massp}\Ap da 
\end{equation}
is unity or higher. The optical depth factors into the calculation of total reflected starlight (Eq.~(\ref{eq:ftot})), describing the attenuation of starlight from scattering by material interior to each radial location on the disk. The result is
\begin{equation}
    f_{\text{total}} = \int_{\ain}^{\aout} \frac{dA}{da} \frac{e^{-\tau_\text{opt}}}{4\pi a^2} da, 
\end{equation}
where, in our disk models,
\begin{equation}\label{eq:dAda}
    \frac{dA}{da} \approx 4 \pi a \Sigma/\massp \Ap
    = \frac{3 \pi \Sin a}{2\rho_p \rratfac \rmin} \left(\frac{a}{\ain}\right)^{-1}.
\end{equation}
The optical depth $\tau_\text{opt}$ is an indicator of how much of the disk is illuminated by the host star. For optically thin disks, $\tau_\text{opt} \ll 1$, all dust particles reflect starlight. For an optically thick disk, we expect reflected starlight only from the inner portion of the disk. The radial width of this region may be inferred by finding the annular width that corresponds to $\tau_\text{opt} = 1$: 
\begin{eqnarray}
\Delta a & = &\ain \times \left[ \frac{3\Sin}{8 h \rratfac\rhop\rmin} - 1\right]^{-1} \\
& = & \ain \times \left[ \frac{3\mdisk}{16 \pi h \ain (\aout-\ain)\rhop\rratfac\rmin} - 1\right]^{-1}
\end{eqnarray}
When the disk is geometrically thin ($h\rightarrow 0$), or when the surface density is large, this annulus becomes a narrow ring at the disk's inner edge. Otherwise, $\Delta a$ can be large, indicating an optically thin disk. Formally negative values indicate that the disk will be optically thin even if its radial extent is infinite.

These assessments of optical depth enable an estimate of how massive a disk can be  before it to remain optically thin. Setting $\tau_\text{opt}=1$ in Equation~(\ref{eq:tauopt} and solving for disk mass, we obtain
\begin{equation}\label{eq:appxmthresh}
M_\tau \sim \frac{32\pi}{3} \ain \aout h \rhop \rratfac \rmin.
 \end{equation}
This threshold mass decreases whenever the density within the disk increases, as when the scale height is reduced, or if the radial extent of the disk as lessened.

\section{Light curve filtering and sensitivity}\label{appx:sensitivity}

The Kepler stars generally have flux levels that are variable even during periods of quiescence. Stellar rotation, coupled with the presence of starspots, causes significant modulation in the light curve --- the active stars in this study are certainly no exception. Yet characterization and ``flattening'' of the quiescent light is essential to echo detection. Fortunately, modulation of the starlight by rotation takes place on time scales of days or weeks, whereas echoes in disks at Kuiper-belt distances span hours.  Thus, a judiciously chosen flatten algorithm, with features of a high-pass filter, can remove the slow modulations. 

However, this flattening process comes with risks. The removal of unwanted low frequency signal amounts to subtracting off a smoothed version of the original signal, generated by a (perhaps) weighted average of time bins in some window (range of bins). The pitfalls are well-known; if the window sharply eliminates some frequencies to rid some unwanted signal, it triggers a ringing around sharp edges in the light curve, notably imperfectly masked flares. The cure is to make the window less sharp in the frequency domain, but doing so will admit the unwanted signal to some degree. The art is to find the right window to jointly minimize the artifacts and the unwanted signal.

\begin{figure}
    \centerline{\includegraphics[width=0.7\textwidth]{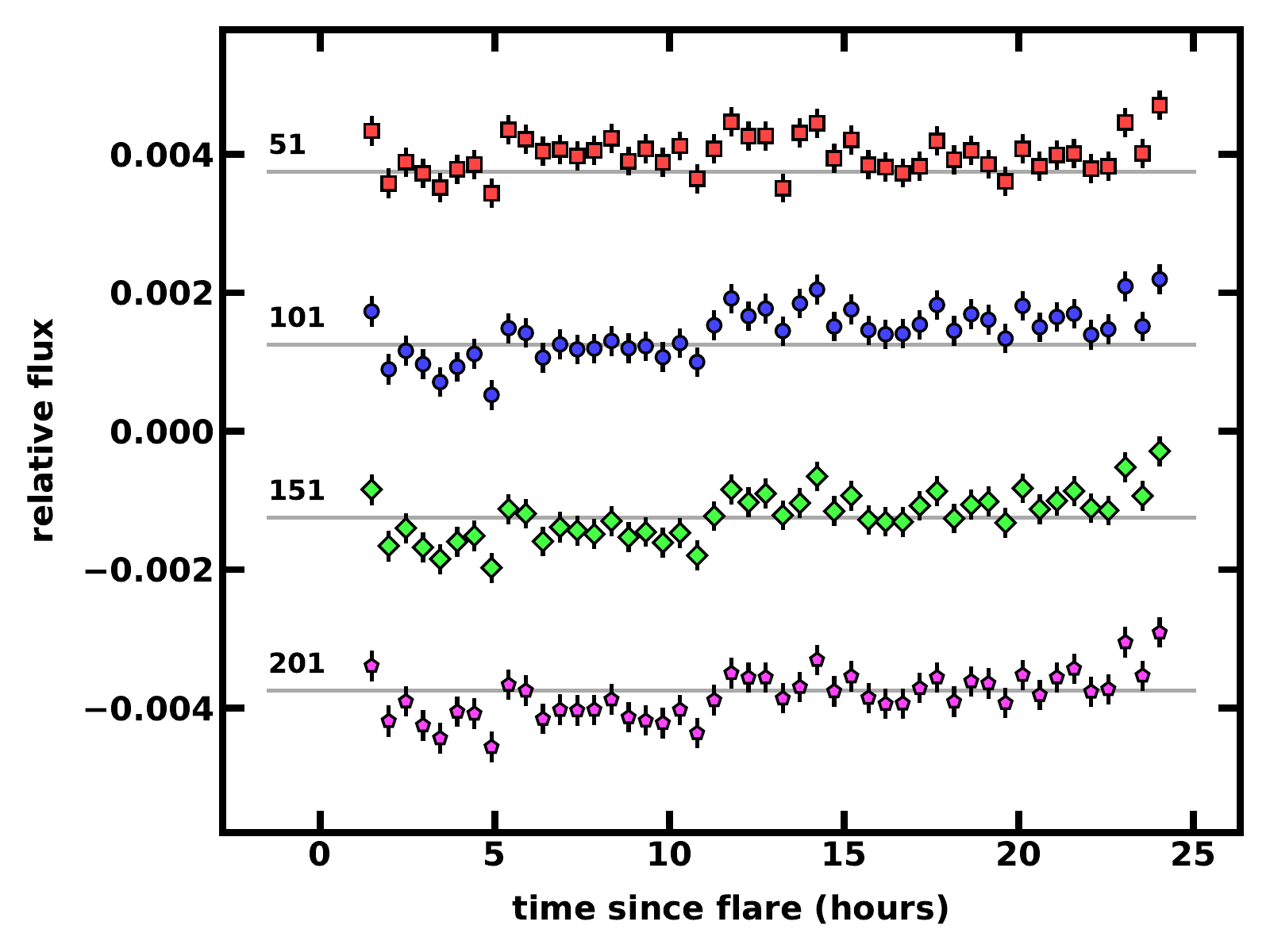}}
\caption{\label{fig:sensitivity}
    Catalog-averaged composite post-flare light curves with variations in the smoothing window size. The points are mean flux values, and error bars give 1-$\sigma$ standard deviation of the mean within each time bin. The curves in each panel have smoothing window sizes (number of time bins is the smoothing window) as labeled; the points in blue,show the smoothing window size adopted here, with 101 bins. 
    }
\end{figure}

The flattening procedure adopted here (\S\ref{subsec:meth}) uses an adaptive Savitzky–Golay filter to limit slowly varying components of the light curve, iterated with steps that mask impulse flares to reduce ringing. The window size (number of bins in the filter) determines the low-frequency cut-off. To assess how our results depend on the choice of windowing, we vary the window size and compare outcomes. Figure~\ref{fig:sensitivity} is an illustration, showing the overall sensitivity of the light curves and the response to variations in window size. The figure contains the catalog-averaged light curves for the long-cadence and short-cadence data, illustrating that the light curves are not strongly dependent on the specific choice of window size.  Within a single light curve, flares that were on the cusp of being added to analysis or rejected as outliers would sometimes end up with different fates for different windows---because no conclusions were drawn from individual light curves, this was not particularly relevant, but a detailed sensitivity analysis would be required if adequate signal-to-noise was available to attempt a detection claim.

When the number of flare events is large, as in a catalog average or long-term observations of a single active star, a flattened baseline may also emerge through simple averaging of light curves. If light-curve modulations from star spots and stellar rotation enter at a typical level of 0.1\% over the duration of post-flare light curves, then roughly $10^6$ flare events are needed to keep relative errors below $10^{-5}$. This strategy could provide a good alternative to filtering in future observational campaigns.

\section{Flare light-curve structure and temporal resolution}\label{appx:unresolved}

\begin{figure}
    \centering
    \includegraphics[width=0.8\textwidth]{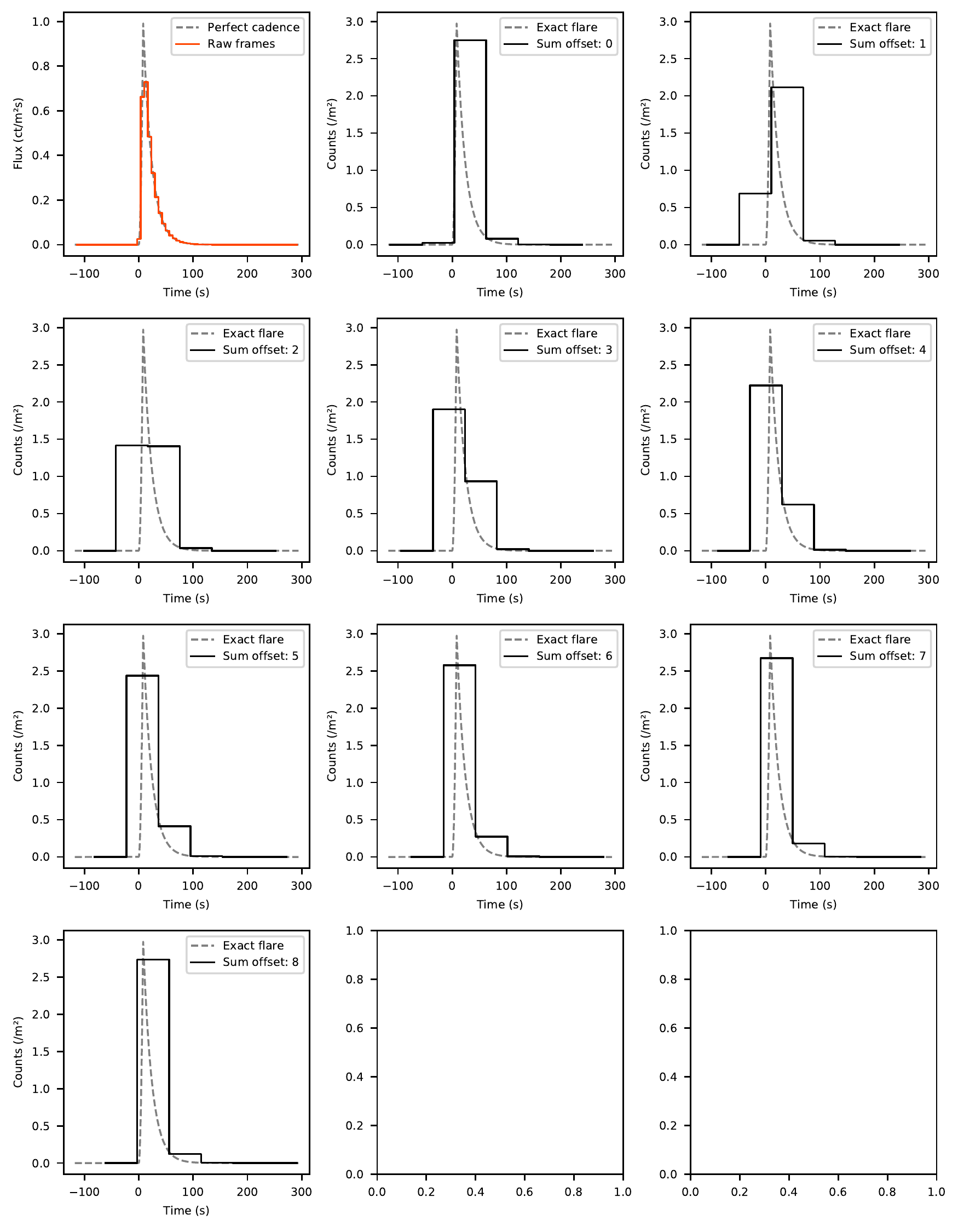}
\caption{\label{fig:cadence_shape}
    Simulation showing impact of frame summation and digitization on perceived flare shape for impulsive flares in short cadence data.  In all cases, the same 'true' flare is measured and the same integration and readout times are used, but the 9-frame summation is shifted by a single frame.  The structure evolves based on the timing of the flare and the percent of light lost to readout peaks around 8\%. Model parameters: parabolic-rise, exponential decay (PRED) flare with a rise time of 8~s, decay constant of 16~s, integration time of 6.0198~s, readout time of 0.519~s, 9-frame sum.  
    }
\end{figure}

As described in \ref{subsec:meth}, the light received is a complex convolution of instrument measurement, the flare signal, and the echo.  Here, we provide a slightly more nuanced version, though we still ignore some critical processing, like PSF integration.  The measured, discretized flux in time bucket $i$ in the vicinity of a flare is:
\begin{equation}
I_i = v(Q(t_i) + \Fflare(t_i) + (\Fflare*G)(t_i))
\end{equation}
where $v$ is the instrument measurement function (quantum efficiency, readout time, integration time, frame stacking or summing, etc.).  For a single frame with no summation, 
\begin{equation}
    I_i = \int_{t_i}^{t_{i+1}-\delta} Q(t)+\Fflare(t)+(\Fflare*G)(t) \,dt
\end{equation}
where $\delta$ is the readout time where no flux is counted.  On Kepler, $\delta= 0.520~s$ which is short relative to a transit but can be somewhat long relative to a flare peak (each exposure is $6.019~s$, so short impulsive flares can lose on the order of ~10{\%} of their light if their timing is unlucky--this is considerable when attempting to perform a measurement that requires sub-percent accuracy).  The summation and discretization process can also impact the measured shape of barely resolved flares.  For example, if a flare occurs part-way between time bins, it is possible to have the integrated peak of the flare appear in a bin adjacent to the actual flux peak, or not appear as a peak at all, as shown by the model used on Kepler short-cadence data in Figure~\ref{fig:cadence_shape}. 

The actual observation is then corrupted by noise, so the vector of measured data is more likely:
\begin{equation}
    x_i=Poisson(I_i)+N_{read,i}
\end{equation}
with $N_{read,i}$ the readout noise in that measurement.  For simplicity, we will  write $x_i=I_i+N_i=Q_i+{\Fflare}_i+(F*G)_i+N_i$.  The goal is to find $G$ or otherwise constrain its existence to some bounding values.  First we run the data through high-pass filter $H$:
\begin{equation}
    \vec{x}_{filt}=H(\vec{x})
\end{equation}
If $H$ is linear and data well-behaved, then we get:
\begin{equation}
    \vec{x}_{filt}=H(\vec{Q}) + H(\vec{F}_{flare}) + H(<\Fflare*G>) + H(\vec{N}),
\end{equation}
where the angular braces denote a temporal average.
Because the noise in each bin is assumed independent and high-frequency, and the background is low frequency, we assume the noise passes through and $Q$ is outright eliminated to arrive at:
\begin{equation}
    \vec{x}_{filt}\sim H(\vec{F}_{flare}) + H(<\Fflare*G>) + \vec{N}.
\end{equation}
This form makes it clear that smooth, extended echoes could be filtered out along with the background if we are not careful.  If the star has superior background stability, it is possible to mask the post-flare region 
replacing it with (e.g.) a polynomial approximation or period-folded template. 
Because each flare is presumed somewhat unique, we can no longer simply sum the flares and preserve the echo structure.  We need to somehow use our knowledge of the measured flare shape for subsequent analysis, such as deconvolution or forward modeling.  

To estimate the actual flare shape before discretization and noise, some assumptions must be made.  Real flares are, on an appropriately fast timescale, smooth and continuous, though they may be quite sharp near the peak.  A good model should not over-soften the peak, so a piece-wise models is often appropriate; 
we adopt such a flare model with a parabolic rise and exponential decay (PRED), with functional form $v_\text{PRED}(\Bar{b}, t)$, where $\Bar{b}$ are the numerical coefficients that estimate the profile's shape.
These coefficients can be found with:
\begin{equation}
    \Bar{b}_{opt}=argmin(||H(v_\text{PRED}(\Bar{b}, t)) - \vec{x}_{filt}||).
\end{equation}
In practice, an option for extracting the optimal coefficients is to attempt to deconvolve each flare from the post-flare light curve using this model, but for an echo buried in noise this approach will have dubious value. 
Another option is to use the flare model to predict the resulting echo and subsequently perform a maximum likelihood analysis.

We now have a format that allows us to test different disk structure hypotheses. For disk model $G_{\Bar{a}}$, we would expect the echo in the processed light curve to look like:
\begin{equation}
\vec{y}_{\Bar{a},\Bar{b}} = H(v_{\text{PRED}} (\Bar{b}_{opt}) * G_{\Bar{a}})
\end{equation}
Each flare in the catalog will have a different $\Bar{b}$, but the same $\Bar{a}$, excepting for an unknown weight function that accounts for which parts of the disk are illuminated, depending on the flare location on the star.  Three options here are to (1) fit this weight function for each flare, (2) use a single composite weight function on each flare, fitting each flare imperfectly intentionally, or (3) combine the estimates and compare them to an average weight function prediction for $G_{\Bar{a}}$.  In the simple models introduced in \S\ref{sec:echoes}, we took this last approach.

For Case 1, fitting phase to each flare likely only makes sense if there is a secondary source of information (e.g., the star has a clear activity cycle correlated with rotational phase, so we can reasonably guess the phase of the active region) or after confirming a likely echo through a coarser means; otherwise, it risks over-fitting the data.  To use this approach, we derive a residual,
\begin{equation}
    R_{j,\Bar{a}}=||\vec{y}_{\Bar{a},\Bar{b}_j}-\vec{x}_{j,filt}||.
\end{equation}
The null hypothesis, no echo, is $ R_{j,0}=||\vec{x}_{j,filt}||$.  For a perfect fit, $R_{j,\Bar{a}}=||\vec{N}||=\sigma_j$ and these cases can therefore be readily compared across the flare catalog with a chi-square test.  

For Case 2, fitting each flare echo to a composite phase function will never fit each flare exactly, but on average will fit.  In that case, $G_{\Bar{a}}$ should be replaced with one integrated over all phases, $\Bar{G}_{\Bar{a}}$, which is represented by $\vec{y}'$, and tested for goodness of fit in each flare, $j$, such as:
\begin{equation}
    R_{j,\Bar{a}}=||\vec{y}'_{\Bar{a},\Bar{b}_j}-\vec{x}_{j,filt}||
\end{equation}
In this case, there will always be some distance between the exact solution and the mean solution, $||\vec{y}'_{\Bar{a},\Bar{b}_j}-\vec{y}_{\Bar{a},\Bar{b}_j}+N||$, with a magnitude that will depend on the parameters being tested.  Therefore, $R_{j,\Bar{a}}$ will have a value that can be estimated for a given disk model. 

For Case 3, the estimates from each flare are combined, such as:
\begin{equation}
    \vec{Y}_{\Bar{a}}=\sum_{j=1} w_j \vec{y}_{\Bar{a},\Bar{b}_j}
    \quad\text{and}\quad 
    \vec{X}_{filt}=\sum_{j=1} w_j \vec{x}_{j,filt}
\end{equation}
with $w_j$ a normalized flare quality weight, such as peak intensity and/or inverse decay constant, and can then be tested:
\begin{equation}
     R_{\Bar{a}}=||\vec{Y}_{a}-\vec{X}_{filt}||
\end{equation}
In this case, the null hypothesis is $ R_{0}=||\vec{X}_{filt}||$. For a sufficiently large number of flares, the solution will converge towards a perfect fit so $R_{\Bar{a}}$ should approach the weighted standard deviation of the mean and these mean cases can again be compared with a Chi-square test.

\bibliography{stellarecho}{}

\end{document}